\documentclass[a4paper, 11pt, oneside]{article}
\usepackage{float}
\usepackage{jheppub}
\usepackage{epstopdf}

\usepackage{multirow}
\usepackage[caption=false]{subfig}

\newlength{\absize}
\def\lsim{\mathrel{\raise.3ex\hbox{$<$\kern-.75em\lower1ex\hbox{$\sim$}}}}
\def\gsim{\mathrel{\raise.3ex\hbox{$>$\kern-.75em\lower1ex\hbox{$\sim$}}}}
\allowdisplaybreaks 

\def\mpt{{\slash\!\!\!\!\!\:P}_T}

\def\mptv{{\slash\!\!\!\!\!\:\vec{P}}_T}

\usepackage{rotating}
\usepackage{verbatim}

\def\i11{{\mathbbm 1}}

\title{Charged-Higgs production in the Two-Higgs-Doublet Model --- 
the $\tau\nu$ channel}

\author[a]{Lorenzo~Basso}
\affiliation[a]{Institut Pluridisciplinaire Hubert Curien/D\'epartement
    Recherches Subatomiques,\\ Universit\'e de Strasbourg/CNRS-IN2P3,
    23 rue du Loess, F-67037 Strasbourg, France}

\author[b]{\!\!, Per~Osland}
\affiliation[b]{Department of Physics and Technology, University of Bergen, \\
Postboks 7803, N-5020  Bergen, Norway}

\author[c]{and Giovanni~Marco~Pruna}
\affiliation[c]{Paul Scherrer Institut, CH-5232 Villigen PSI, Switzerland}

\emailAdd{Lorenzo.Basso@iphc.cnrs.fr}
\emailAdd{Per.Osland@ift.uib.no}
\emailAdd{Giovanni-Marco.Pruna@psi.ch}
\date{\today}

\abstract{We update the allowed parameter space of the CP-violating 2HDM with Type~II Yukawa couplings, that survives the current experimental and theoretical constraints on the model. For a representative set of allowed parameter points, we study the production of charged Higgs bosons, both at the LHC at 14~TeV and at a possible future hadronic collider at 30~TeV. Two classes of production mechanisms are considered, ``bosonic'' ($pp\to H^\pm W^\mp X$) and ``fermionic'' ($pp \to H^+ \bar t (b) X$). After commenting on our previous $H^\pm\to W^\pm H_1$ investigation, we focus on 
the tauonic decay mode, $H^\pm \to \tau\nu$, performing a detailed signal-over-background analysis at the parton level. The increased features provided when considering CP violation, i.e., the extension of the parameter space and the mixing of the would-be CP-odd scalar boson, only marginally increase the discovery prospects, which remain very challenging both when increased luminosities and higher energies are considered.}


\begin{document}
\begin{flushright}
PSI-PR-15-04 \\
\end{flushright}
\maketitle


\section{Introduction}
\label{sec:intro}
After the discovery of the Higgs boson \cite{Aad:2012tfa,Chatrchyan:2012ufa}, the major experimental challenges concerning the scalar sector of the Standard Model (SM) are pointing in two directions: on the one hand, there is a general interest in the accurate determination of the Higgs couplings in order to establish the exact nature of the particle and possible deviations from the standard scenario; on the other hand, a tireless search for other scalar resonances is conducted in order to possibly reveal the non-minimality of the Higgs sector.

Focusing on the latter, a special case is represented by the search for a charged Higgs boson. Indeed, such particle would reveal not only the presence of Beyond the SM (BSM) physics, but also a scenario that goes beyond minimal scalar singlet extensions. From this perspective, charged Higgs searches are widely considered a central part of new-physics (NP) searches. 

One of the most popular realisations of a theory containing a charged Higgs boson is the so-called Two-Higgs-doublet model (2HDM), since it can also be taken as representative for manifestations of the Higgs sector of a supersymmetric (SUSY) framework at the electro-weak (EW) scale, when the SUSY spectrum is decoupled from the SM. Assuming that SUSY particles lie outside the LHC reach, in the absence (so far) of any SUSY signal, the 2HDM setup corresponds to a rather motivated phenomenological model. In its more general construction, the additional doublet also provides more CP violation \cite{Lee:1973iz} than the usual SM one, induced by the CKM matrix only. This feature is especially welcome for baryogenesis \cite{Riotto:1999yt}, and it comes accompanied with a wider and phenomenologically richer parameter space.

Concerning the Yukawa sector, there are different schemes for introducing it in the 2HDM, referred to as type~I, type~II, type~X (often labelled type~III), or type~Y (type~IV). Depending on the Yukawa couplings, different structures of the interactions are involved and, as a consequence, different experimental constraints apply. We shall here be interested in the type~II model, where one doublet (here referred to as $\Phi_2$) couples to up-type quarks, and the other doublet ($\Phi_1$) couples to down-type quarks, as well as to the charged leptons. This is the same structure as that of the Minimal Supersymmetric Standard Model (MSSM), and historically this type has therefore received more attention.

The ``disadvantage'' of this scenario is that the Yukawa couplings are such that charged-Higgs exchange would contribute to the process
\begin{equation}\label{eq:btoxsgamma}
\bar B\to X_s\gamma,
\end{equation}
for which there is excellent agreement with the Standard Model (SM), where the transition is mediated only by $W$ exchange. The result is that the charged-Higgs mass is severely constrained, and a lower bound of about $380$~GeV has to be imposed \cite{Hermann:2012fc}. Usually, for lower allowed masses, the dominant production channel is the one connected to $t$-quarks produced in the initial state, further decaying in $H^\pm+X$. However, when the aforementioned lower mass bound is imposed, the overall scenario is certainly more intriguing, as there is neither a preferential production nor decay channel.

For $m_{H^\pm}\gtrsim 400$ GeV, it was recently shown \cite{Basso:2012st,Basso:2013hs,Basso:2013wna} that the channel
\begin{equation} \label{Eq:WH1-channel}
H^\pm\to W^\pm H_1,
\end{equation}
where $H_1$ is the SM-like Higgs, leading to the overall chain
\begin{equation}\label{discovery}
pp \to H^\pm W^\mp X \to W^+ W^- H_1 X \to jj\ell\nu b\bar b X,
\end{equation}
can be detected in the Run 2 of the LHC experiments for a considerable region of the non-excluded CP-violating (CPV) 2HDM type~II parameter space. This mode was also studied recently for the CP-conserving case \cite{Coleppa:2014cca}. In that case, there are two channels corresponding to (\ref{Eq:WH1-channel}), namely 
\begin{equation}
H^\pm\to W^\pm H/W^\pm A,
\end{equation}
where $H$ is the heavier CP-even and $A$ the CP-odd Higgs boson. In the alignment limit (see the next section and in particular, Eq.~(\ref{Eq:alignment})), there is no such coupling to the lightest CP-even Higgs boson, $h$.

Among the much-explored decay channels, a particular relevance is generally devoted to the tau channel:
\begin{equation}
H^\pm \to \tau^\pm \stackrel{(-)}{\nu}.
\end{equation}
This is due to its cleaner nature with respect to the quark counterpart $H^\pm\to tb$ and to its importance in determining the leptonic Yukawa sector in the most accurate way, the tau being the heaviest among the leptons.

In this paper, first the parameter space of the CPV 2HDM type~II is updated, then the channel in Eq.~(\ref{discovery}) is briefly reanalysed to confirm its discovery potential at the LHC at Run~2. Subsequently, possible strategies for detecting a charged Higgs decaying into the leptonic third generation at present and future hadronic colliders are described.

The paper is organised as follows. In section~\ref{Sec:2} we review the model. In section~\ref{Sec:3} we present an overview of the viable parameter space, subject to theoretical and experimental constraints. The phenomenological study of the model is the central core of the paper. In particular, the various signals are discussed in section~\ref{Sec:4}, while in section~\ref{Sec:5} we review the backgrounds and present the result of our signal-over-background investigation. Section~\ref{sec:conclusions} contains our conclusions, and an appendix presents a quantitative discussion of box-diagram contributions.
A brief summary of preliminary results was presented in Ref.~\cite{Basso:2014npa}.

\section{Model}
\label{Sec:2}
\setcounter{equation}{0}
The most common and simplest version of the 2HDM potential is here considered, similarly to the previous study of \cite{Basso:2012st}, i.e., without terms
proportional to $\lambda_6$ and $\lambda_7$. Such terms would lead to
flavour-violating neutral interactions at the tree level,
which are severely constrained \cite{Adam:2013mnn,Aubert:2009ag}. In Feynman gauge, the two Higgs doublets are decomposed as
\begin{equation}
\Phi_i=\left(
\begin{array}{c}\varphi_i^+\\ (v_i+\eta_i+i\chi_i)/\sqrt{2}
\end{array}\right), \quad
i=1,2.
\label{Eq:basis}
\end{equation}
The neutral sector comprises
3 scalars, $H_j$ ($j=1,2,3$), not restricted to CP eigenstates, which are defined through the diagonalisation
of the mass-squared matrix, ${\cal M}^2$, by an orthogonal rotation matrix $R$:
\begin{equation} \label{Eq:R-def}
\begin{pmatrix}
H_1 \\ H_2 \\ H_3
\end{pmatrix}
=R
\begin{pmatrix}
\eta_1 \\ \eta_2 \\ \eta_3
\end{pmatrix},
\end{equation}
satisfying
\begin{equation}
\label{Eq:cal-M}
R{\cal M}^2R^{\rm T}={\cal M}^2_{\rm diag}={\rm diag}(M_1^2,M_2^2,M_3^2).
\end{equation}
The rotation matrix $R$ is parametrised in terms of three angles,
$\alpha_1$, $\alpha_2$ and $\alpha_3$ \cite{Accomando:2006ga,Basso:2012st}.
In Eq.~(\ref{Eq:R-def}), $\eta_3=-\sin\beta\chi_1+\cos\beta\chi_2$, orthogonal to the neutral Goldstone boson. The charged Higgs boson is defined by the same rotation:
\begin{equation}
H^\pm=-\sin\beta\varphi_1^\pm+\cos\beta\varphi_2^\pm,
\end{equation}
and $\tan\beta=v_2/v_1$.

In this study, the $H_j H^\mp W^\pm$ coupling plays an important
role. In the CP-violating model, with all momenta incoming, it is given by \cite{El_Kaffas:2006nt}
\begin{equation} \label{Eq:HHW}
H_j H^\mp W^\pm: \qquad
\frac{g}{2}
[\pm i(\sin\beta R_{j1}-\cos\beta R_{j2})+ R_{j3}]
(p_\mu^j-p_\mu^\mp).
\end{equation}

For the charged Higgs boson, we have for the Yukawa
coupling to the third generation of quarks \cite{Gunion:1989we}
\begin{alignat}{2}  \label{Eq:Yukawa-charged-II}
&H^+ b \bar t: &\qquad
&\frac{ig}{2\sqrt2 \,m_W}\,V_{tb}
[m_b(1+\gamma_5)\tan\beta+m_t(1-\gamma_5)\cot\beta], \nonumber \\
&H^-  t\bar b: &\qquad
&\frac{ig}{2\sqrt2 \,m_W}\,V_{tb}^*
[m_b(1-\gamma_5)\tan\beta+m_t(1+\gamma_5)\cot\beta],
\end{alignat}
and similarly for the coupling to $\tau\nu$, substituting $V_{tb}\to1$, $m_t\to0$ and $m_b\to m_\tau$.

\section{Parameter space}
\label{Sec:3}
\setcounter{equation}{0}
The model parameters are subject to the following constraints:
\begin{itemize}
\item
Theory constraints: positivity, unitarity, global minimum, as
described in our previous paper \cite{Basso:2012st}.
The checking for a global minimum is performed by solving a set of three
coupled cubic equations \cite{Grzadkowski:2010au}.
\item
The low-energy flavour constraints as listed in our previous paper
\cite{Basso:2012st}, including the $S,T,U$ constraints and the
constraint on the (CP-violating) electron electric dipole moment. Penalties for all
these are added in a $\chi^2$ measure, and disallowed parameter points are cut off at $3\,\sigma$.
\item
LHC constraints are treated generously, in view of the frequent
updates of experimental results. The signal strengths
$\mu_{\gamma\gamma}$, $\mu_{ZZ}$ and $\mu_{\tau\tau}$ are evaluated, and
parameter points violating any one of these by more
than $3\,\sigma$ \cite{David:2014,Kado:2014} are excluded. (They are not compounded to an
overall $\chi^2$, since we have no quantitative information on the correlations.) The couplings of $H_2$ and $H_3$ to $WW$ are evaluated, and only parameter points
corresponding to non-discovery \cite{CMS:2012bea,Chatrchyan:2013yoa,TheATLAScollaboration:2013zha,Aad:2013dza} of such heavier states are kept.
\end{itemize}

Subject to these constraints, and with ``physical'' input in terms of
mass parameters and mixing angles as described elsewhere
\cite{Khater:2003wq}, we
sample selected discrete values of $\tan\beta$, $M_2$, $M_{H^\pm}$, and $\mu$,
each with a scan over 5 million trial sets of mixing angles,
$\{\alpha_1,\alpha_2,\alpha_3\}$.
With this input, and with $\lambda_6=\lambda_7=0$, the heaviest mass,
$M_3$, is a derived quantity.

Allowed regions in the $\alpha$ space were presented earlier
\cite{Basso:2012st,Basso:2013wna}. The most recent updates on
$\mu_{\gamma\gamma}$ and $\mu_{ZZ}$, as well as the heavy-Higgs
exclusions \cite{CMS:2012bea,Chatrchyan:2013yoa,TheATLAScollaboration:2013zha,Aad:2013dza}, constrain these further.

The $H_j H^\mp W^\pm$ coupling (\ref{Eq:HHW}) is involved in the production of $H^\pm$ via an intermediate $H_2$ or $H_3$ in the $s$-channel, and it is involved in the decay $H^\pm\to W^\pm H_1$ that we studied previously \cite{Basso:2012st}.
The factor in the square bracket of Eq.~(\ref{Eq:HHW}) can be written as
\begin{alignat}{2}
j&=1: &\quad 
&\pm i\cos\alpha_2\sin(\beta - \alpha_1) + \sin\alpha_2, \\
j&=2: &\quad 
&\mp i[\sin\alpha_2\sin\alpha_3\sin(\beta - \alpha_1) +\cos\alpha_3\cos(\beta-\alpha_1)]
+ \cos\alpha_2\sin\alpha_3, 
\label{eq:H2H_chW}\\
j&=3: &\quad 
&\pm i[-\sin\alpha_2\cos\alpha_3\sin(\beta-\alpha_1) + \sin\alpha_3\cos(\beta-\alpha_1)]
+ \cos\alpha_2\cos\alpha_3. \label{eq:H3H_chW}
\end{alignat}

In the alignment limit, which is closely approached by the LHC data,
with $H_1$ even under CP and with the $H_1ZZ$ coupling like in the SM,
we would have \cite{Grzadkowski:2013rza}
\begin{equation} \label{Eq:alignment}
\beta=\alpha_1, \quad \alpha_2=0.
\end{equation}
Thus, the $H_1H^\pm W^\mp$-coupling vanishes, whereas the absolute
values squared of the above expressions become unity for both $H_2$
and $H_3$. We note that this is in accord with the familiar
CP-conserving alignment limit
\cite{Gunion:1989we}, both the $HH^\mp W^\pm$ and $AH^\mp W^\pm$
couplings have full strength, whereas the $hH^\mp W^\pm$ coupling
vanishes.

\begin{figure}[htb]
\includegraphics[width=\linewidth]{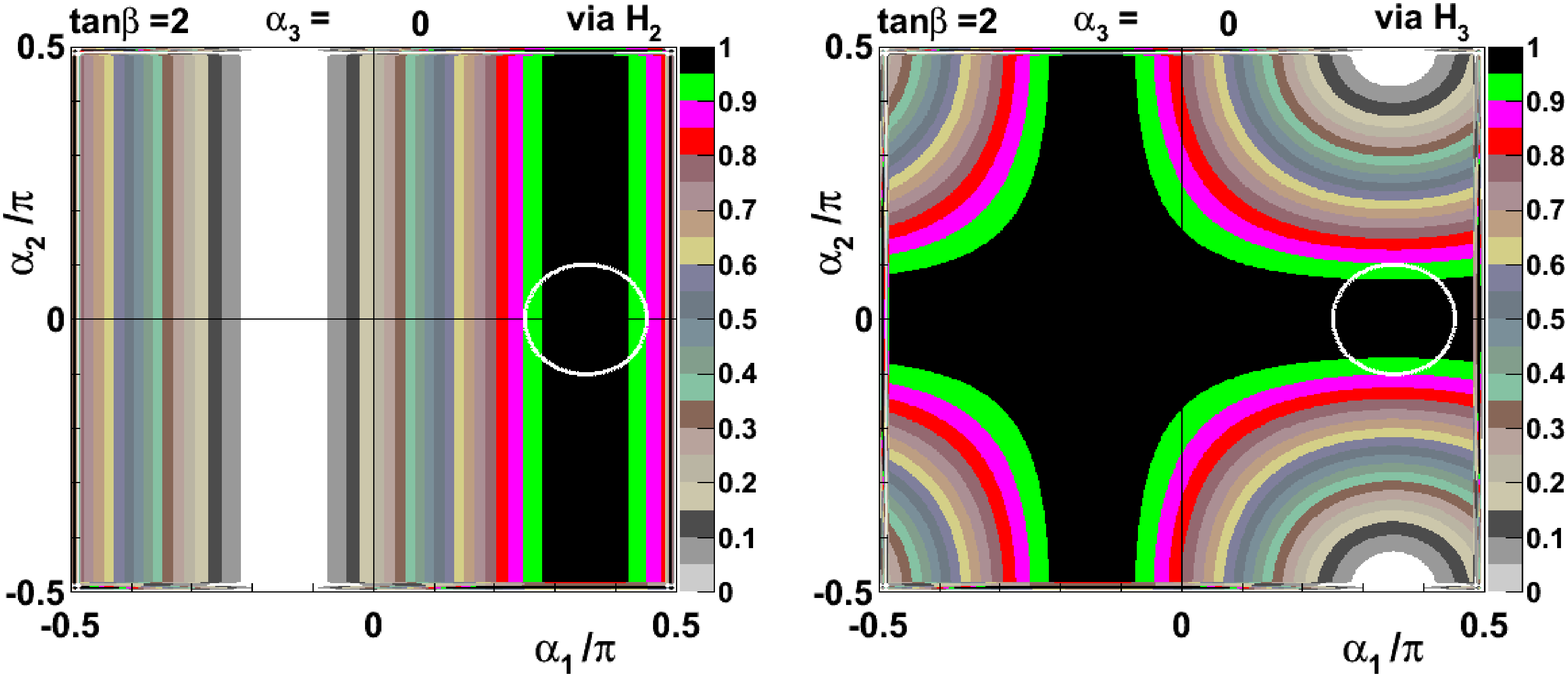}
\includegraphics[width=\linewidth]{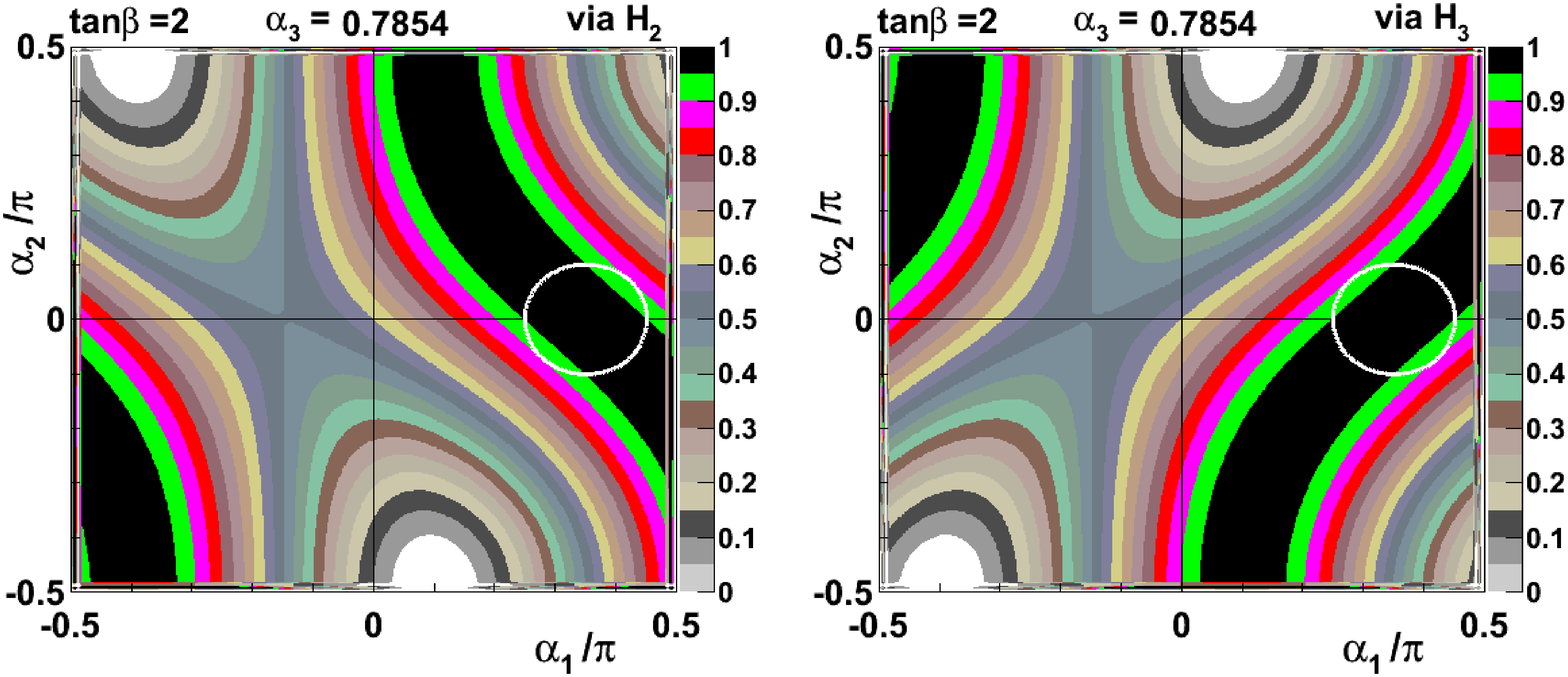}
\caption{Relative rates for $H_2\to H^\pm W^\mp$(left panel) and
  $H_3\to H^\pm W^\mp$(right). These are shown as fractions of the
  maximal rates (for given masses) in the $\alpha_1-\alpha_2$ plane,
  for $\tan\beta=2$. Top:  $\alpha_3=0$: bottom: $\alpha_3=\pi/4$. 
The white circle identifies the region of alignment.
\label{fig:rate-tanbeta02}}
\end{figure}

For $\tan\beta=2$ and two values of $\alpha_3$, namely
$\alpha_3=0$ and $\alpha_3=\pi/4$, we show in
figure~\ref{fig:rate-tanbeta02}
the absolute values squared of the expressions (\ref{eq:H2H_chW}) and
(\ref{eq:H3H_chW}). We see that these saturate at unity (shown in
black) in bands including the alignment limit $\alpha_1=\beta$ and
$\alpha_2=0$.  In fact, it is easy to see from Eqs.~(\ref{eq:H2H_chW})
and ~(\ref{eq:H3H_chW}) that near the alignment limit
(\ref{Eq:alignment}) there is no dependence on $\alpha_3$, as
reflected in figure~\ref{fig:rate-tanbeta02}.  The white ``circle'' shows the region
in which the $H_1ZZ$ coupling agrees with that of the SM to better
than 5\%.

We restrict our studies to values of $\tan\beta\leq10$. Beyond this
point, the model becomes very fine-tuned \cite{WahabElKaffas:2007xd},
in order not to violate unitarity
\cite{Kanemura:1993hm,Akeroyd:2000wc,Arhrib:2000is,Ginzburg:2003fe,Ginzburg:2005dt}.

\section{Phenomenology}
\label{Sec:4}
\setcounter{equation}{0}
In this section, the phenomenology of the production of the
charged-Higgs boson and its decay in the $\tau\nu_\tau$ mode are
analysed in the context of present and future colliders.
Before presenting cross sections, branching ratios and numbers of
events, we shall introduce some terminology and an overview of the tools used.

\subsection{Terminology}
In hadronic collisions, there are several relevant charged-Higgs production channels.
We shall divide them into two categories, ``bosonic'' and ``fermionic''. At the partonic level,  these concepts will be used as follows:
\begin{itemize}
\item ``(A) bosonic'':\quad
$gg\to H_i\to H^\pm W^\mp$,
\item ``(A) bosonic'':\quad
$qq'\to W^\pm \to H^\pm H_i$,
\item ``(B) fermionic'':\quad
$g\bar{b}\to H^+ \bar{t}+\mbox{charge conjugated}$,
\item  ``(B) fermionic'':\quad
$gg\to H^+ b\bar{t}+\mbox{charge conjugated}$.
\end{itemize}
The second channel in the list, i.e., the off-shell $W$-mediated production, is
sub-dominant in our investigation given the large charged-Higgs mass.
From now on, the treatment
will focus on the other three channels unless otherwise specified.
This distinction of bosonic vs fermionic production will play a central role in our discussion.

Two main experimental scenarios will be considered, to which we
generally refer as ``present'' and ``future'' collider
frameworks. Schematically, with these two labels the following
experimental features are summarised:
\begin{itemize}
\item {\bf present}: hadron collider with $\sqrt{s}=14$ TeV and $L=100$ fb$^{-1}$, according to the Run 2 of the LHC.
\item {\bf future}: hadron collider with $\sqrt{s}=30$ TeV and $L=100$ fb$^{-1}$, according to the hypothetical ``HE-LHC'' prototype \cite{Bruning:2002yh,Todesco:2013cca}.
\end{itemize}
The ``present'' and ``future'' scenarios are defined by their centre-of-mass energies. Possible luminosity upgrades (realising the so-called ``HL-LHC'' prototype, e.g., when $L=1$ ab$^{-1}$) can be retrieved by a trivial rescaling.

\subsection{Tools}
Since we want to study a considerable number of allowed points (as
discussed in Section~\ref{Sec:3}), a certain level of automation is
required. The following publicly available tools were exploited both
for computational purposes and for cross-checks:
\begin{itemize}
\item the Lagrangian of the model was implemented both in {\tt LanHEP v3.1.9}\footnote{The Higgs sector of the model, including $H_i\to gg,\gamma\gamma,\gamma Z$ was implemented in LanHEP according to the description in \cite{Mader:2012pm}, while the Yukawa sector was borrowed from \cite{Basso:2012st}.} \cite{Semenov:2010qt} and in {\tt FeynRules v2.0} \cite{Alloul:2013bka}, and the agreement of the Feynman Rules produced by the two packages was checked;
\item for the study of the box contributions to the $gg\to H^\pm W^\mp$ partonic process, the combined packages {\tt FeynArts v3.9}~\cite{Hahn:2000kx} and {\tt FormCalc v8.3}~\cite{Hahn:1998yk,Nejad:2013ina} were employed. The integrated cross sections (numerically evaluated with the {\tt Collier} library~\cite{Denner:2014gla}) have been cross-checked by the evaluation of the non-integrated amplitudes, symbolically manipulated with {\tt Form v4.0}~\cite{Kuipers:2012rf} and  numerically evaluated with the package {\tt LoopTools 2.10}~\cite{Hahn:1998yk};
\item the calculation of cross sections and branching fractions as well as the generation of events for the signal was done in {\tt CalcHEP v3.4.6}~ \cite{Belyaev:2012qa} with the {\tt CTEQ6L} PDF set~\cite{Pumplin:2002vw}. For the evaluation of the ``bosonic'' signal, only triangle vertices have been implemented. We shall comment on this approximation in Appendix~\ref{App:A};
\item the generation of the background events was performed with {\tt MadGraph5\underline{ }aMC@NLO v2.1.2}~\cite{Alwall:2014hca} employing the {\tt CTEQ6L1} PDF set;
\item the event analysis was done with the {\tt MadAnalysis~5 v.1.1.12} package~\cite{Conte:2012fm,Conte:2014zja}.
\end{itemize}

\subsection{Signal}

In this subsection, an analysis of charged-Higgs-mediated
signals at the LHC is presented. In addition to the charged-Higgs tau
decay mode, we shall also comment on the previously analysed \cite{Basso:2012st,Basso:2013hs,Basso:2013wna} purely bosonic production and decay channel $pp\to H^\pm W^\mp\to W^\pm W^\mp H_1$. In the following, we
discuss the two scenarios that above have been labelled as ``present'' and ``future''.

\begin{figure}[ht]
\includegraphics[width=0.48\linewidth]{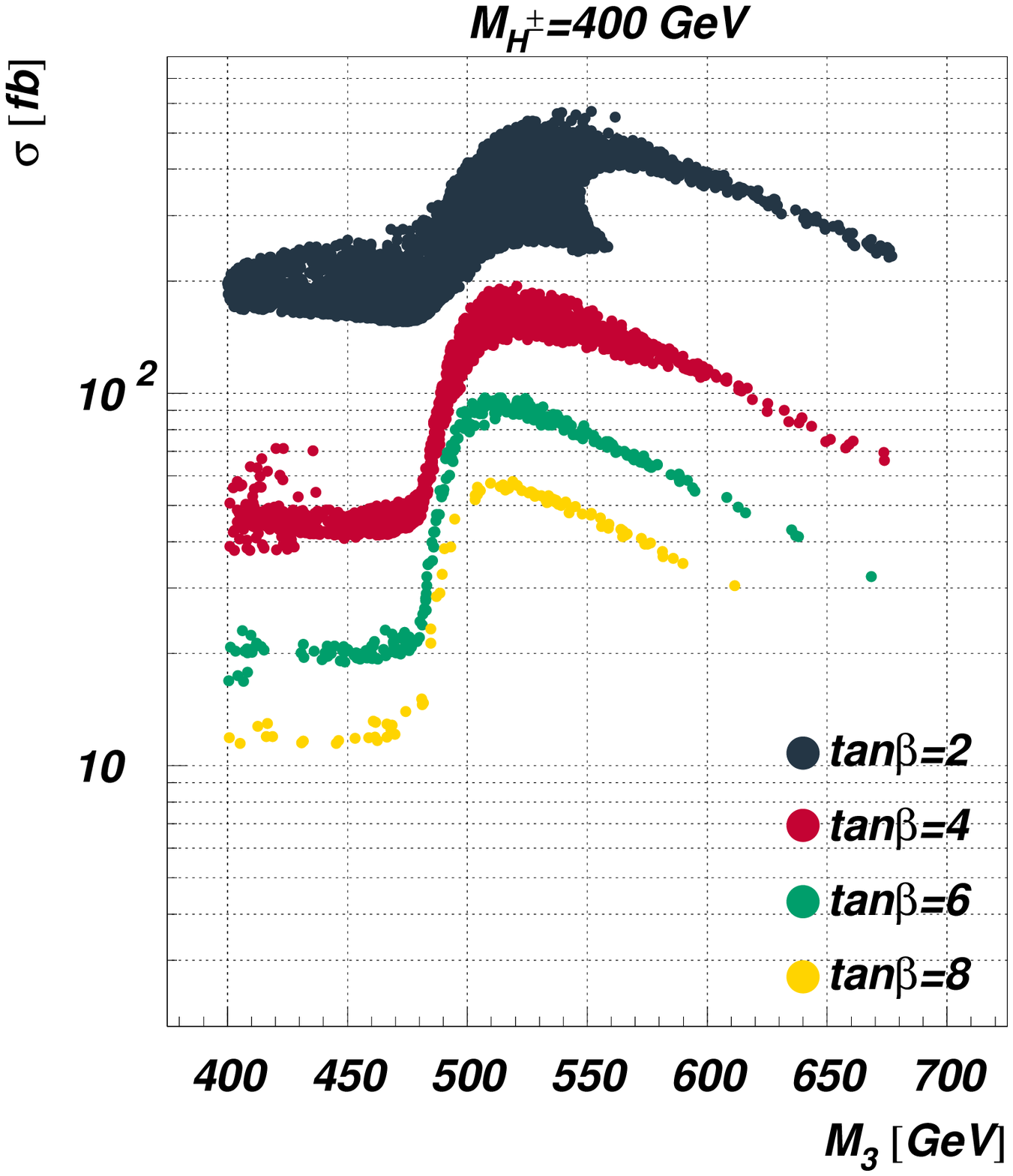}
\includegraphics[width=0.48\linewidth]{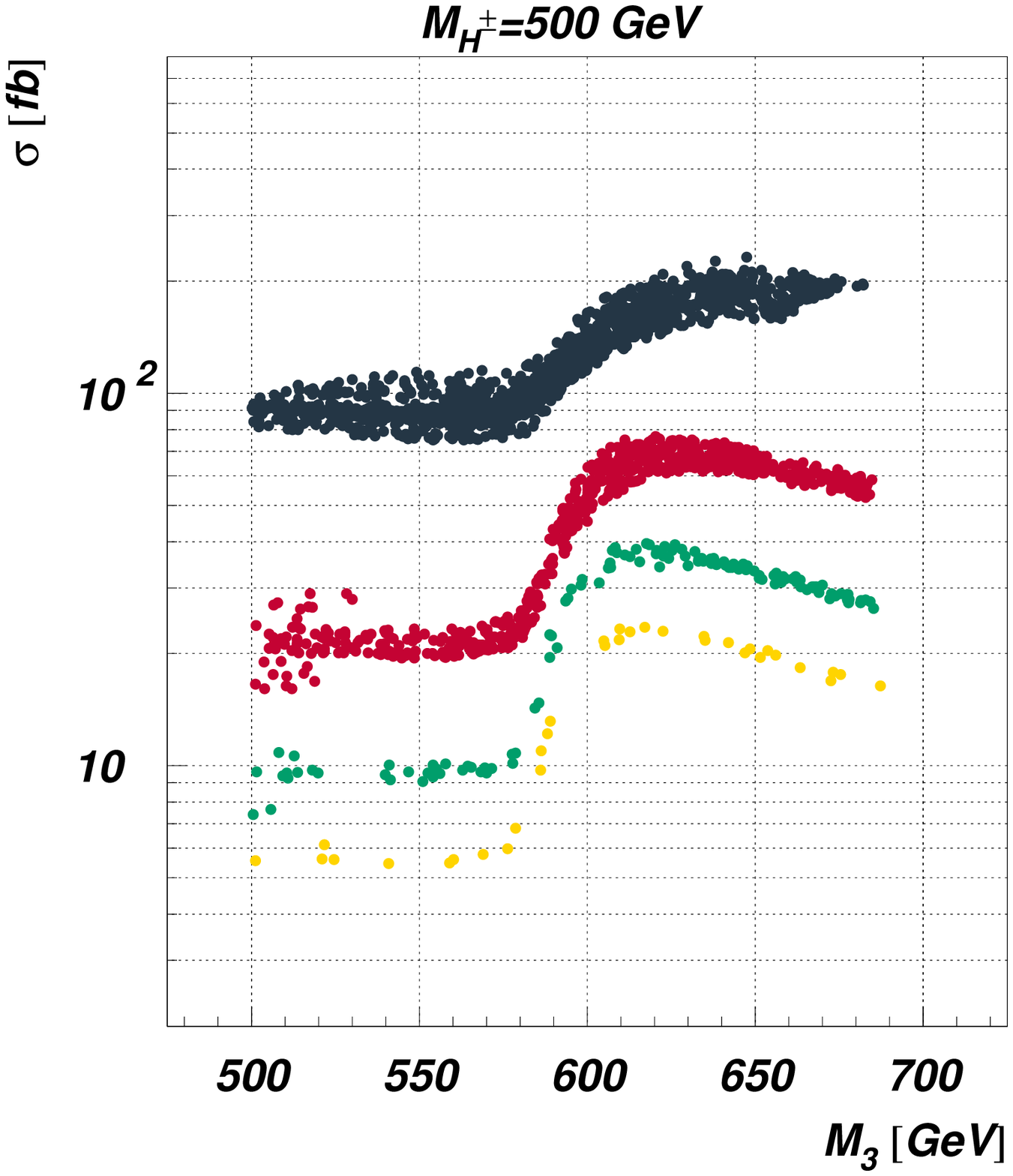} \\
\includegraphics[width=0.48\linewidth]{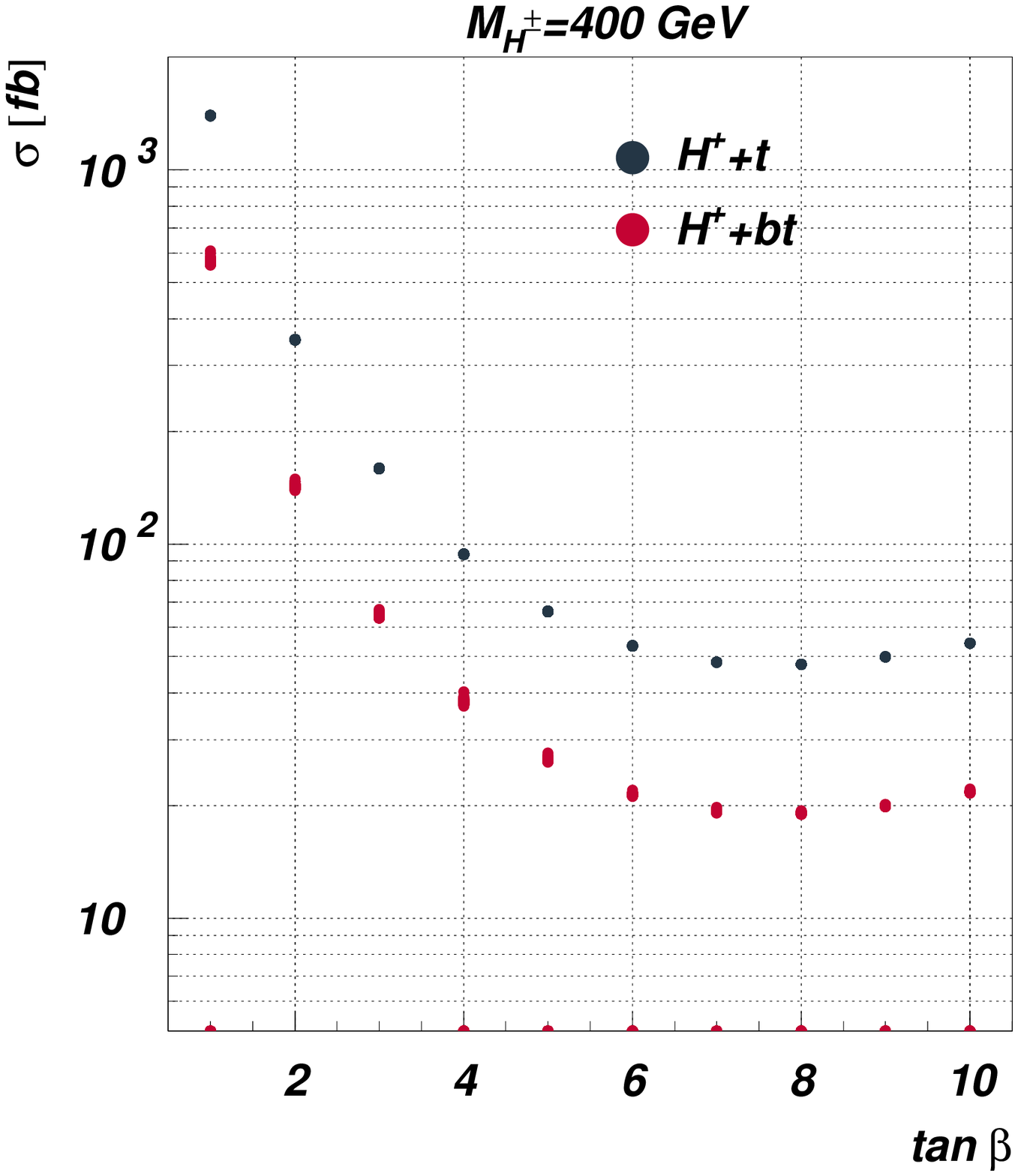}
\includegraphics[width=0.48\linewidth]{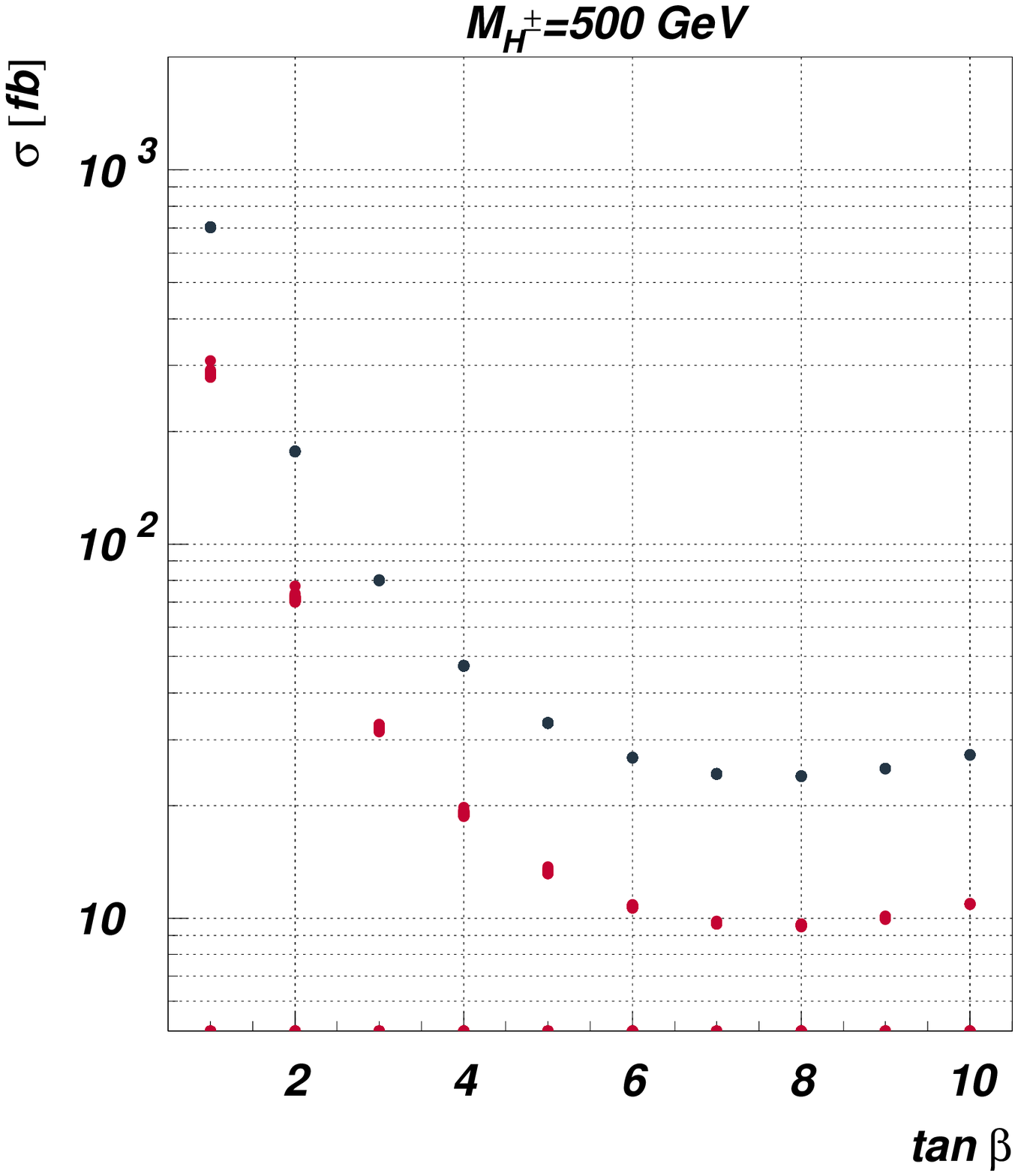}
\caption{Cross sections for the main production channels. Top: bosonic production mode plotted vs $M_3$.
Bottom: fermionic production modes plotted vs $\tan\beta$.
Two charged-Higgs masses are considered, 400 and 500~GeV, at the Run 2 of the LHC. \label{fig:production}}
\end{figure}

In figure~\ref{fig:production}, the cross sections for the main
production channels are plotted against relevant quantities: for the
bosonic case (upper panels), there is a resonant behaviour due to the
presence of a neutral scalar $H_3$, whereas for the fermionic case
(lower panels), the trend is strictly dictated by the value of
$\tan{\beta}$. In both cases, low values of $\tan{\beta}$ lead to an
increased production, while the cross sections drop for higher
values. In the fermionic case, there is a minimum corresponding to the
minimum value of the coupling $H^\pm\to tb$,
i.e. $\tan{\beta}=\sqrt{m_t/m_b}\sim 8$, then the cross section
increases again. Hence, the best scenario for the charged Higgs
\emph{production} occurs in the bosonic case for low values of
$\tan\beta$, and when $M_3\sim M_{H^\pm}+M_W$. The ``bosonic'' cross
sections have been here evaluated in the approximation of considering
only triangle diagrams and neglecting the box ones. 
By doing so, and given the negative interference between triangle and
box diagrams, the bosonic cross sections is overestimated. However,
when the process gets resonant, i.e. for $M_3 > M_{H^\pm}+M_W$, the
relative impact of neglecting the box diagrams gets smaller and
smaller as $M_3$ increases. In the rest of this paper we will focus on
the resonant production, that is the only case where the bosonic process yields cross sections that
can be observed above the background. In this case, as shown in
Appendix~\ref{App:A}, the error of neglecting the box diagrams amounts
to $\mathcal{O}(10\%)$, that is compatible with the parton level
accuracy of our study. Hence, this approximation is justified.
For the fermionic case,
the best scenario occurs for very low or for very high values of $\tan{\beta}$.
 The case with $M_{H^\pm}=500$ GeV reflects the same behaviour as of
$M_{H^\pm}=400$ GeV, with an overall lower production rate due to the
reduced phase space.

\subsubsection{The $H^\pm\to W^\pm H_1$ decay mode}
The above cross-section information must be combined with a study of
the decay modes to better understand the possibilities for a
phenomenological detection. Once the production rates are given, the
subsequent step is to connect them with the analysis
of \cite{Basso:2012st,Basso:2013hs,Basso:2013wna}.

There, the scope of the LHC in exploring the CP-violating 2HDM through
the discovery of a charged Higgs boson produced in association with a
$W$ boson, with the former decaying into the lightest neutral Higgs
boson and a second $W$ state (altogether yielding a $bbWW$ signature)
was considered. Among various sets of surviving points, a few
benchmark points with peculiar behaviours were chosen and a further
event analysis was performed: after the application of standard
detector cuts, the light Higgs and the $W$ boson were reconstructed,
and a top veto was applied. A further strategy to suppress the background
was pursued, that proved to be crucial especially in the case of
the $t\overline{t}$ component. Schematically, it is based on the fact
that signal events will have the distributions of either the invariant mass of 
 $M(b\bar{b}jj)$ or of the transverse mass of 
$M_T(b\bar{b}l\nu)$ that peak around $M_{H^\pm}$, depending on the decay
channel (hadronic or semileptonic, respectively) of the $W$ boson produced by
the charged Higgs, while those stemming from the $t\overline{t}$ background
tend to have distributions that peak around $2 m_t$.
Therefore, when $M_{H^\pm}$ is much greater than $2 m_t$,
it was shown that the background could be significantly suppressed.

\begin{figure}[ht]
\includegraphics[width=0.48\linewidth]{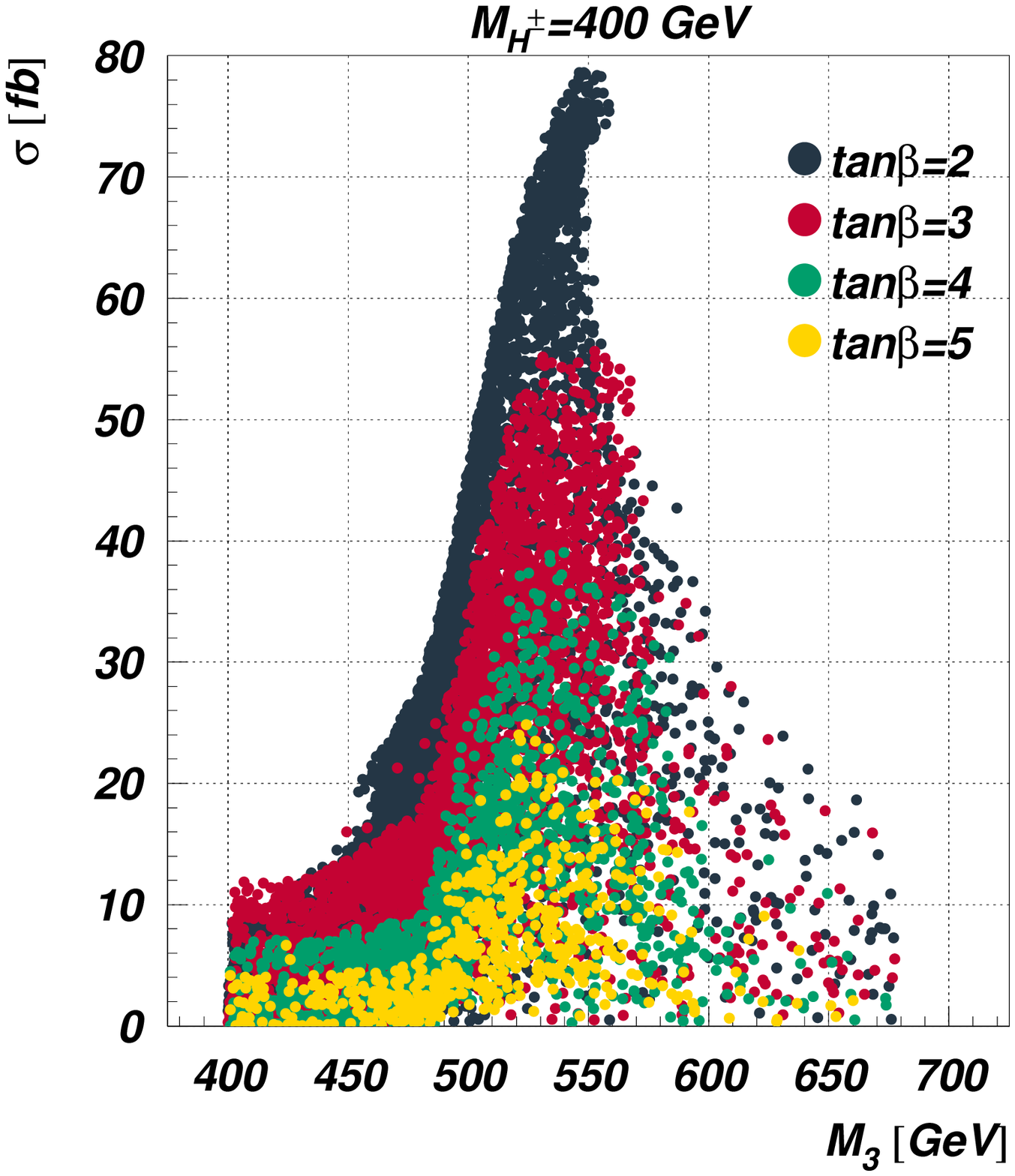}
\includegraphics[width=0.48\linewidth]{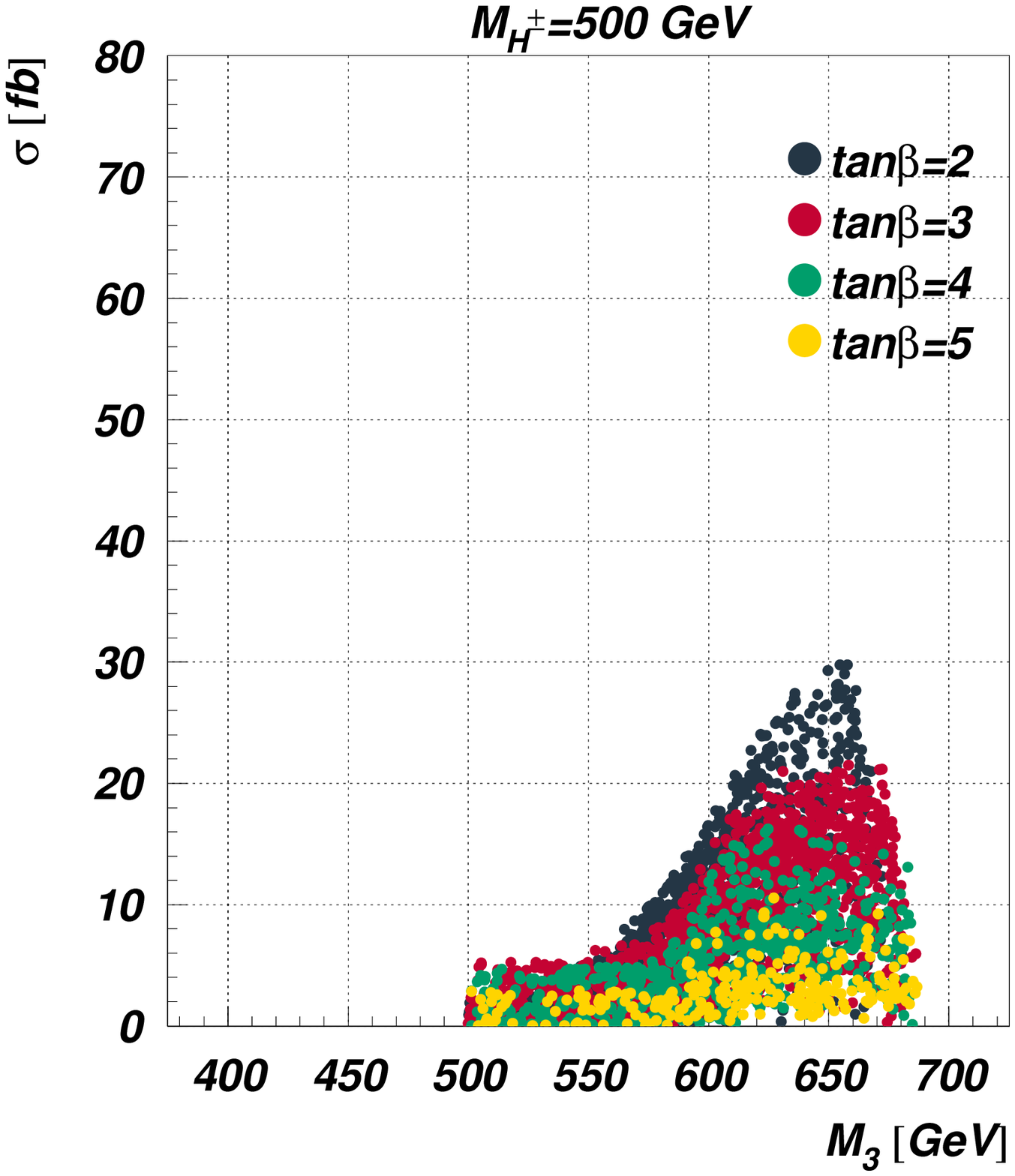}
\caption{Cross section for the $pp\to H_i\to H^\pm W^\mp\to
  W^\pm W^\mp H_1$ channel plotted vs the mass of the heaviest neutral
  scalar $M_3$, for $M_{H^\pm}=400$ (500)~GeV in the left (right)
  panel. Several values of $\tan\beta$ are considered.
\label{fig:discovery}}
\end{figure}

Since we now have
a larger sample of allowed points, as well as updated experimental
constraints, it is of interest to comment on the ``purely bosonic''
production and decay charged-Higgs channel, i.e.
\begin{equation}
 pp\to H_i\to H^\pm W^\mp\to W^\pm W^\mp H_1.
\end{equation}
The production rate associated to this channel is shown in figure~\ref{fig:discovery}. 

After a luminosity of $L=300$ fb$^{-1}$ is collected at the Run 2 of
the LHC, it was previously shown that a cross section of
$\mathcal{O}(50)$ fb is sufficient to extract a signal with a
significance above $\Sigma=3$ for a mass $M_{H^\pm}=400$ GeV. The
proposed method is even more efficient for higher values of the
charged Higgs mass, but a detailed analysis is beyond the scope of
the present paper. 
For the fermionic production mode, a study of this channel 
was published recently \cite{Enberg:2014pua}.

Here, a more general remark is relevant: among the points of the
surviving parameter space, a large number of them remains in the range
where a discovery of the charged Higgs in association with a purely
bosonic production and decay is possible. The favoured region, again,
is for lower values of $\tan{\beta}$, as one can easily infer from
figure~\ref{fig:discovery}.

\begin{figure}[ht]
\centering
\includegraphics[width=0.48\linewidth]{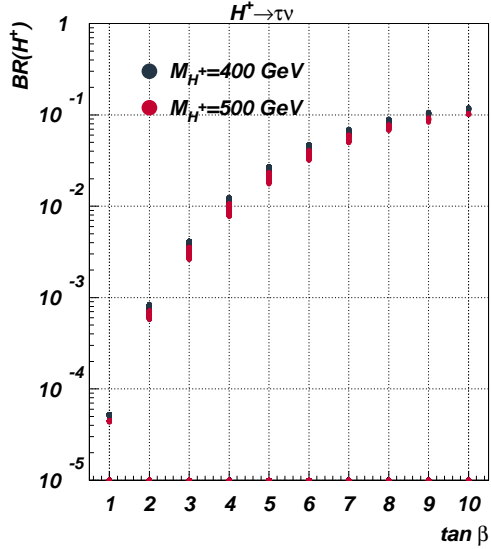}
\caption{Branching ratios of the charged-Higgs $\tau\nu$ decay vs $\tan\beta$. \label{fig:BRHctn}}
\end{figure}

\begin{figure}[ht]
\includegraphics[width=0.48\linewidth]{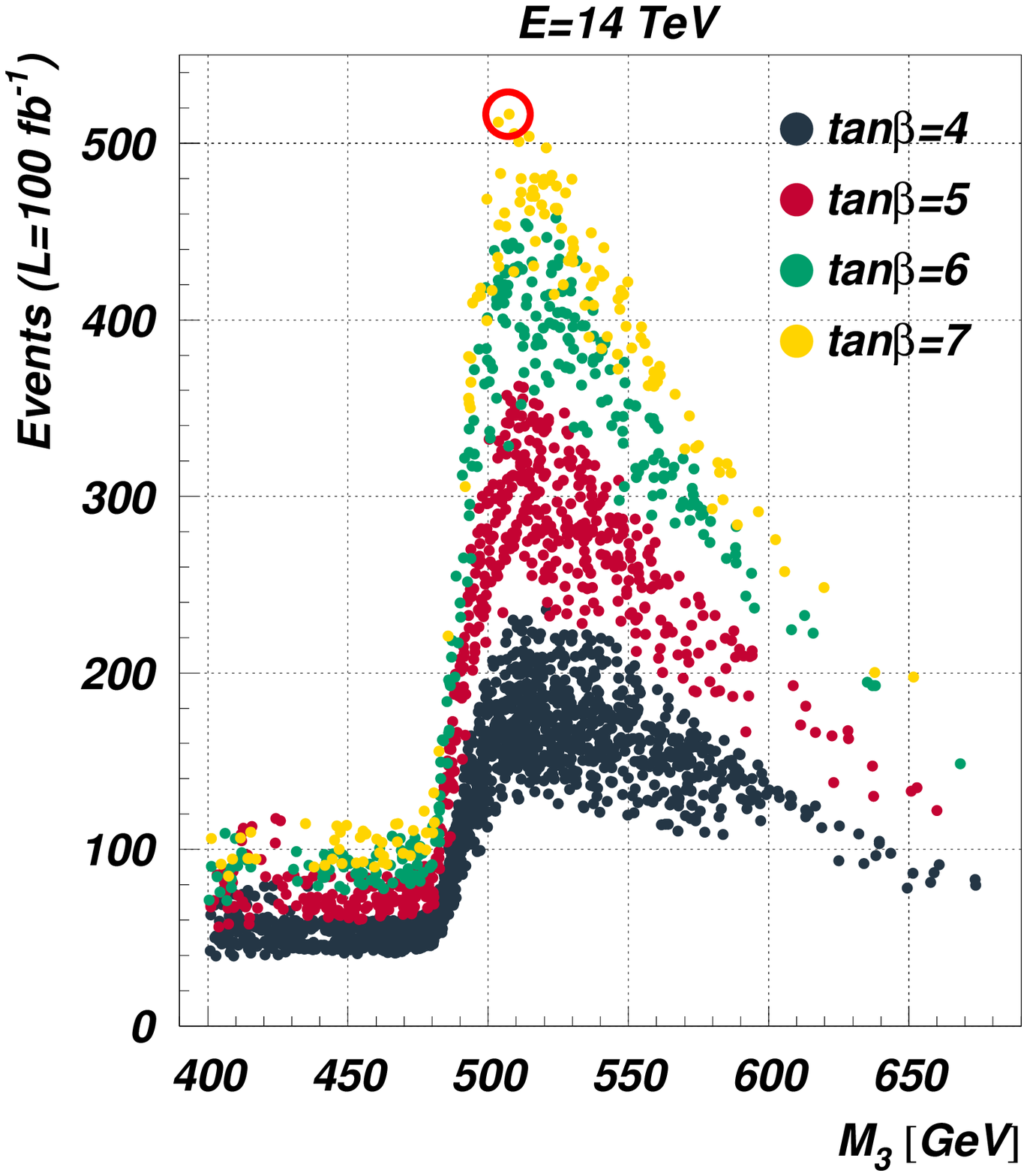}
\includegraphics[width=0.48\linewidth]{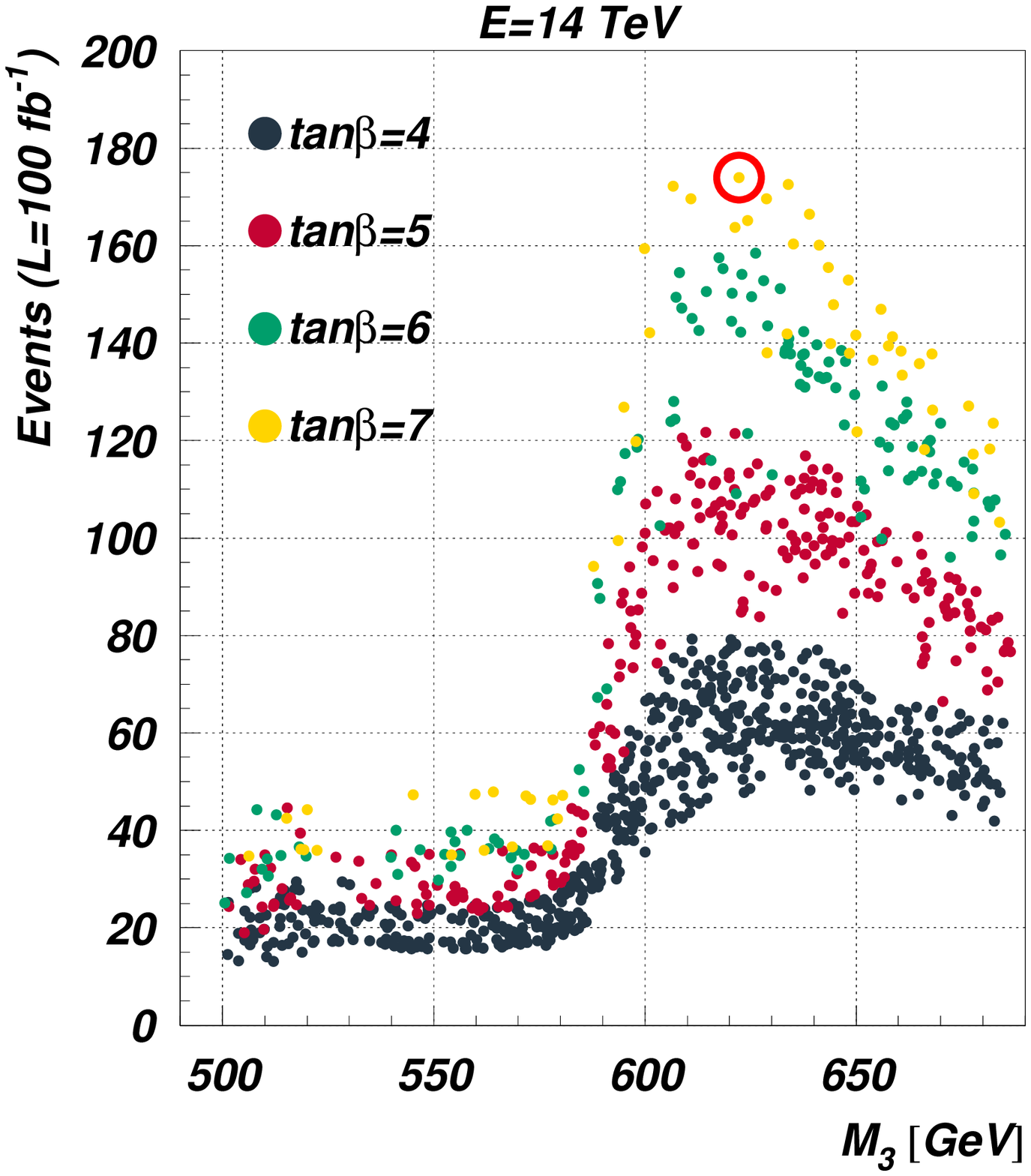}\\
\includegraphics[width=0.48\linewidth]{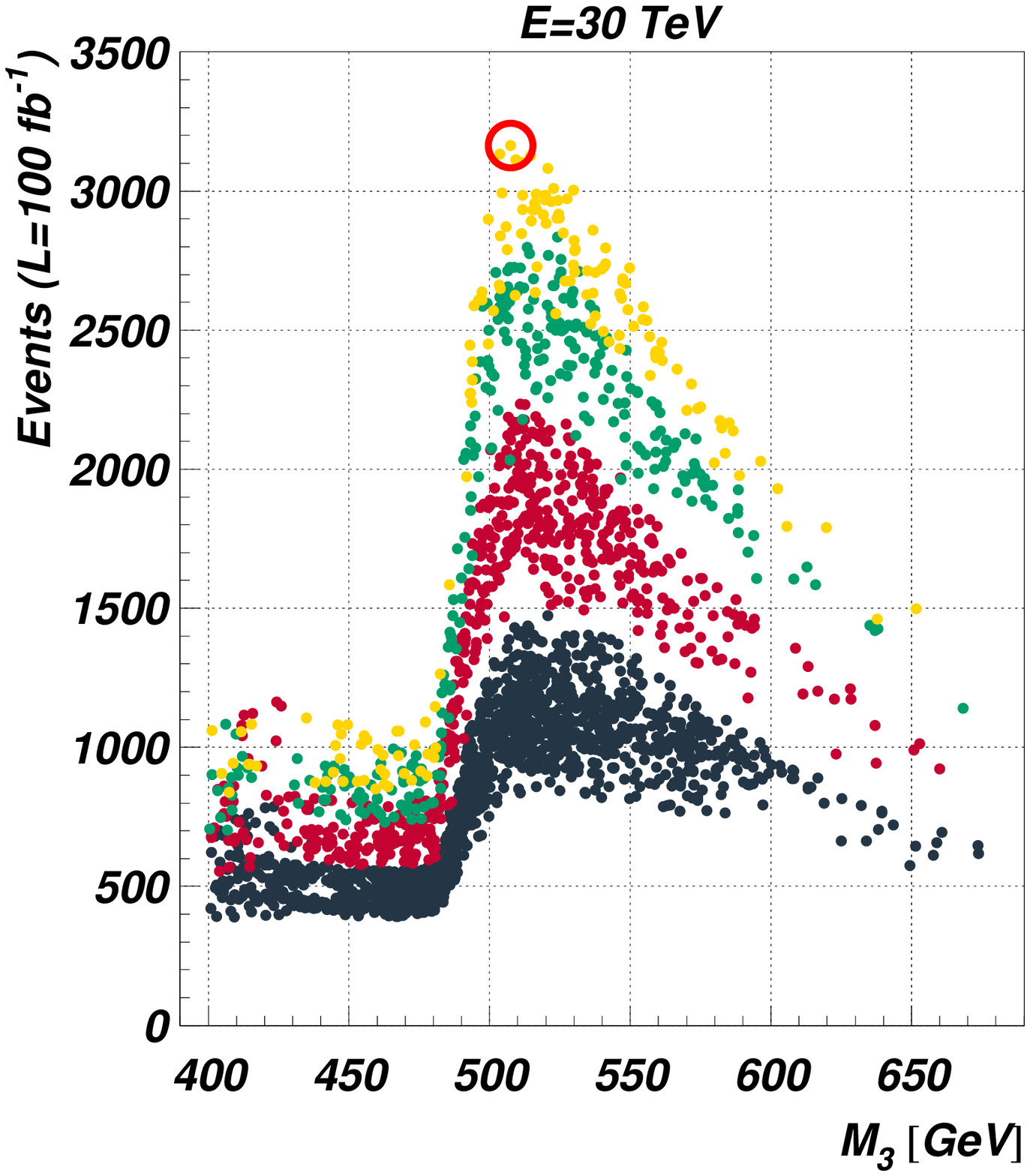}
\includegraphics[width=0.48\linewidth]{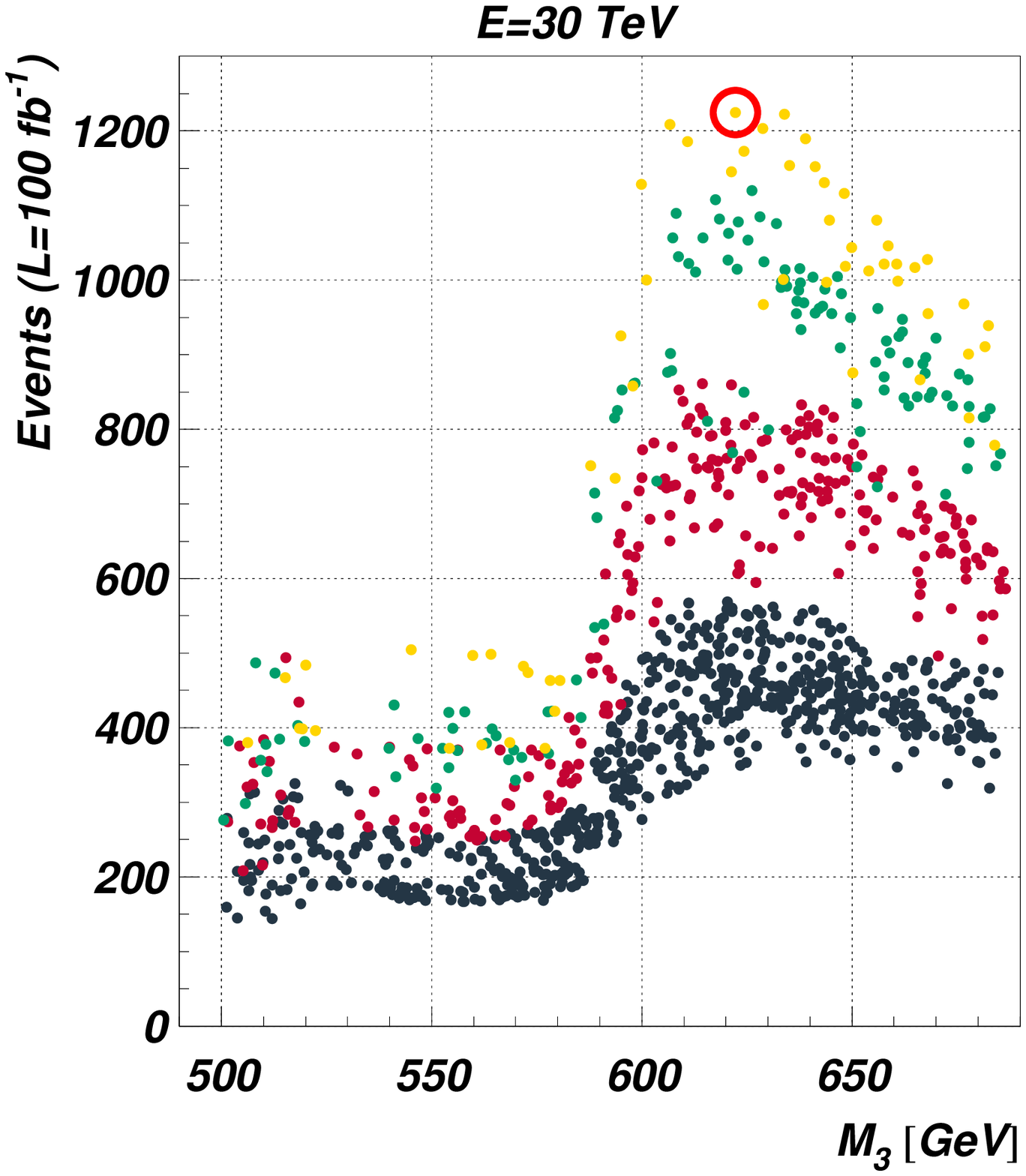}
\caption{Number of events produced via
  $gg\to H_i\to H^\pm W^\mp\to \tau\nu W$ vs $M_3$ at $\sqrt{s}=14$
  (30)~TeV in the upper (lower) panels for various benchmarks with
  $M_{H^\pm}=400$~GeV (left panels) and $M_{H^\pm}=500$~GeV (right
  panels). All are for an integrated luminosity of
  $L=100$~fb$^{-1}$. Red circles indicate the points with the highest
  production rate. \label{fig:MH3_B}}
\end{figure}

\subsubsection{The $H^\pm\to \tau\nu$ decay mode}
The main focus of the present paper is the investigation of the
$H^\pm\to \tau^\pm\nu$ decay modes. In figure~\ref{fig:BRHctn}, the BR
of the charged Higgs tauonic decay is plotted against $\tan{\beta}$,
which again is the only relevant parameters to be considered.

Unlike the cross section, the trend is here reversed: low values of
$\tan{\beta}$ strongly disfavour such a decay mode, that instead
becomes more and more important as $\tan{\beta}$
increases.\footnote{We did not explore values of $\tan\beta$ beyond
  10, since the model then becomes very fine-tuned in order to
  accommodate the unitarity constraints.} This feature yields an
intriguing scenario: the production cross section and the $\tau\nu$
branching ratio are mutually in conflict with respect to the value of
$\tan{\beta}$, only the combined study of these two would finally
reveal the region of the parameter space with highest phenomenological
impact.

In figure~\ref{fig:MH3_B}, the number of events for the bosonic
charged-Higgs production channel with a subsequent charged-Higgs
$\tau\nu$ decay are plotted against the heaviest neutral scalar mass $M_3$ both for a ``present'' and ``future'' scenario.

\begin{figure}[ht]
\includegraphics[width=0.48\linewidth]{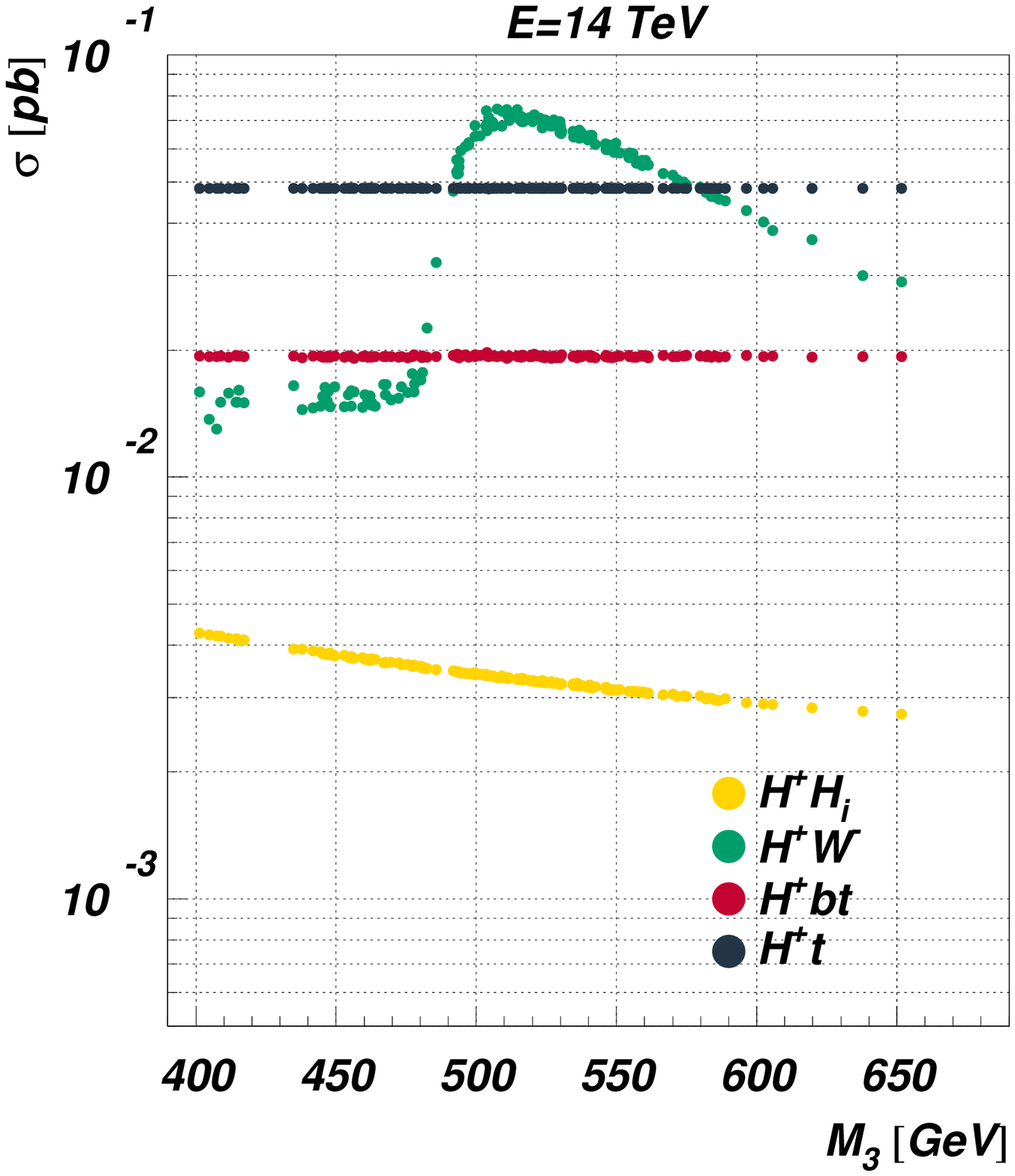}
\includegraphics[width=0.48\linewidth]{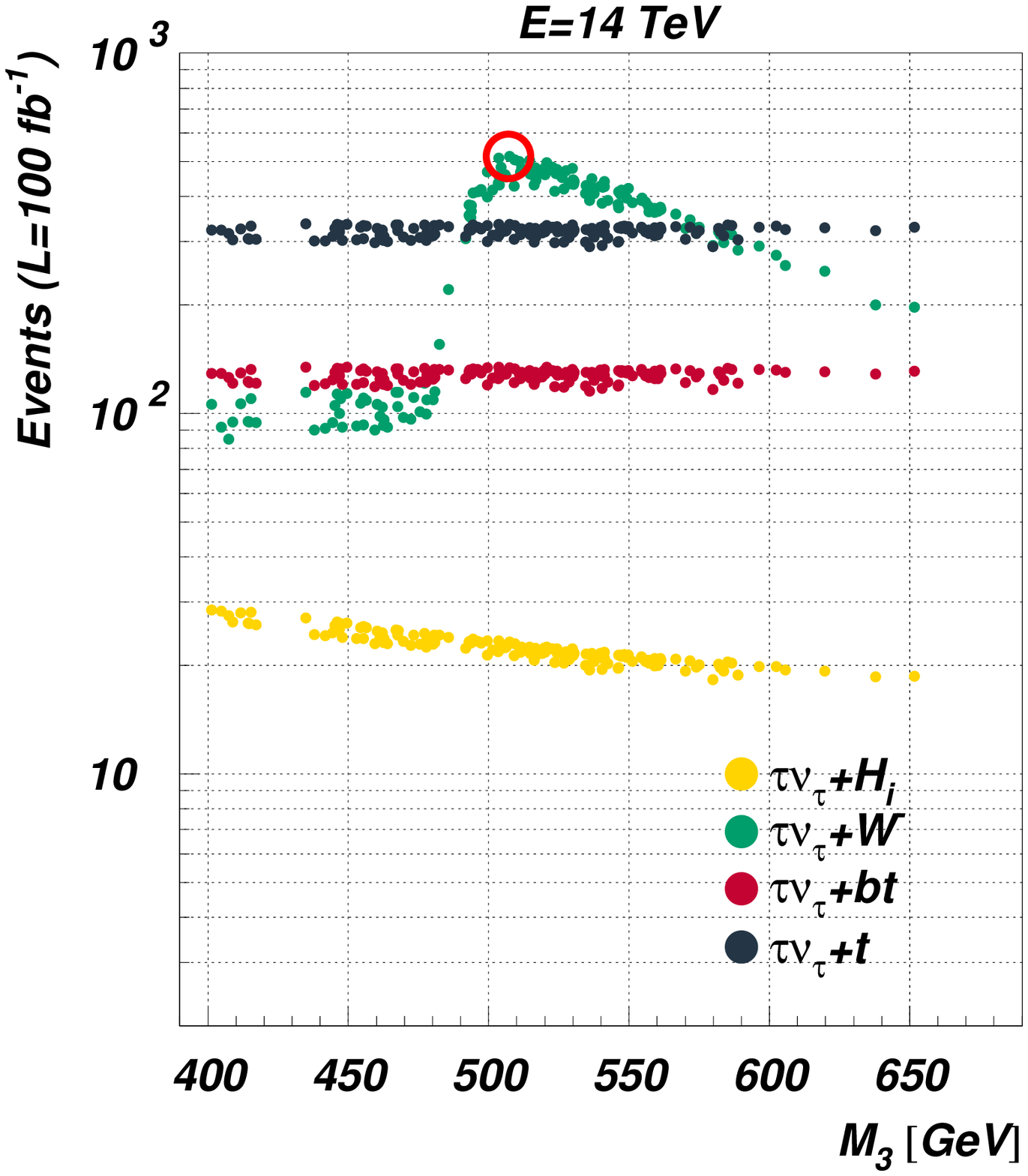}
\caption{Charged-Higgs production cross sections at the Run 2 of the
  LHC (left) and related $\tau\nu+X$ events with $L=100$~fb$^{-1}$
  (right) vs $M_3$. Here, $M_{H^\pm}=400~\text{GeV}$ and
  $\tan\beta=7$. The red circle indicates the points with the highest
  production rate. \label{fig:TB_F}}
\end{figure}

Considering the bosonic production, its combination with the tauonic
decay leads to a situation in which the overall channel is
favoured around $\tan{\beta}\sim 7$--$8$. Among such points, those with highest 
rates are identified by red circles in
the plots. In order to understand what is happening for the
benchmarks around $\tan\beta=7$ (e.g. for a choice of $M_{H^\pm}=400$
GeV), in figure~\ref{fig:TB_F} both the charged-Higgs production
cross sections (left panel) and the number of final-state events in the
``present'' scenario (right panel) are plotted against $M_3$.
By weighting the plot in the left panel by the BRs of
figure~\ref{fig:BRHctn}, and then scaling them by the considered
luminosity, one gets the plot in the right panel. Here, the remarkable
result is that when the intermediate $H_3$ boson is produced resonantly then
the cross section of the bosonic channel is overwhelming with
respect to the one of the fermionic channel. In order to understand
if such behaviour is peculiar of this specific realisation of the
2HDM, a set of benchmark points for the CP-conserving case\footnote{We
  considered the case of $\alpha_2=\alpha_3=0$, when $H_3$ is odd
  under CP.} was produced. In all the studies performed for the
CP-conserving case, the fermionic channel always gives the
highest production rate.

\begin{figure}[ht]
\includegraphics[width=0.48\linewidth]{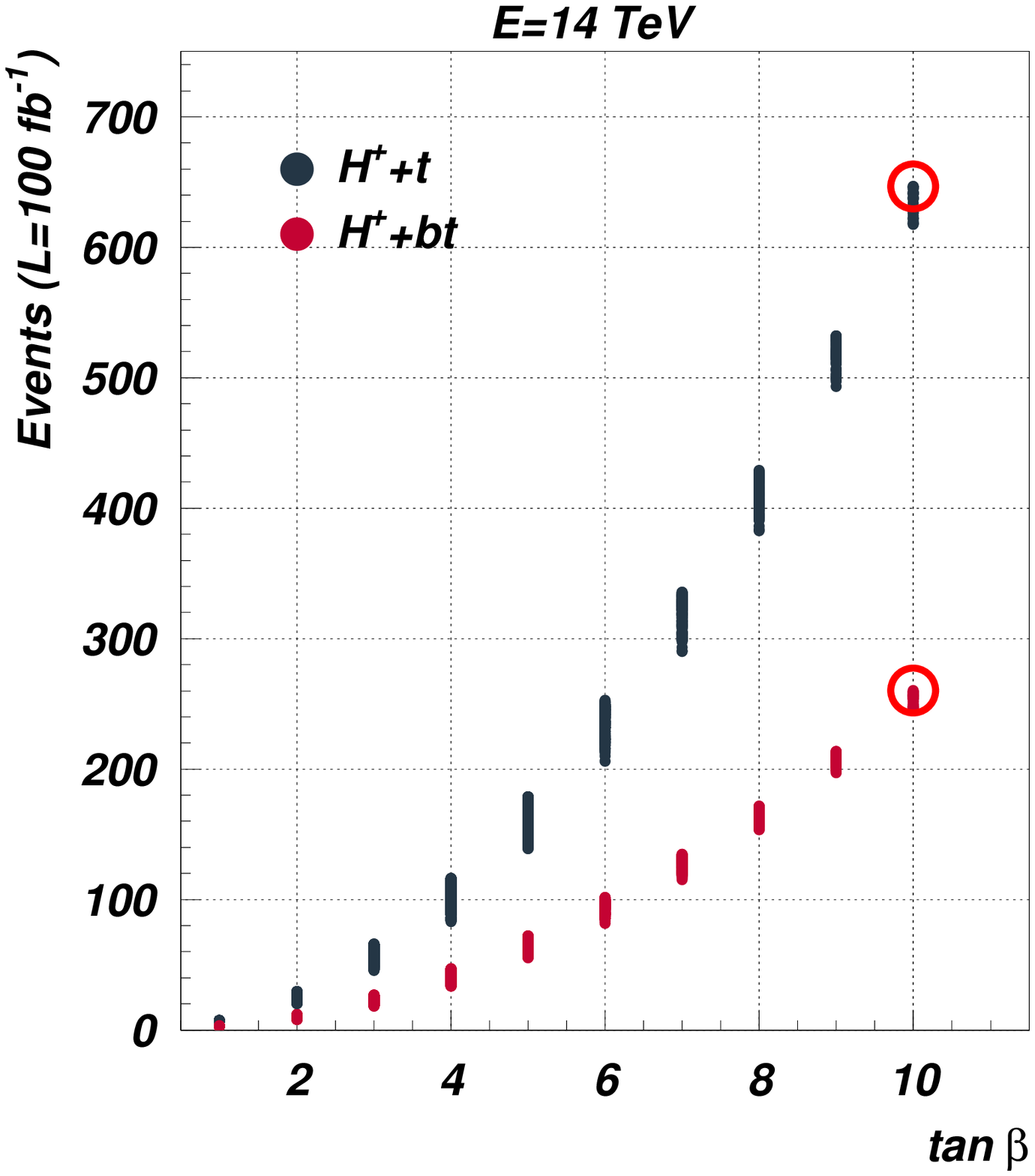}
\includegraphics[width=0.48\linewidth]{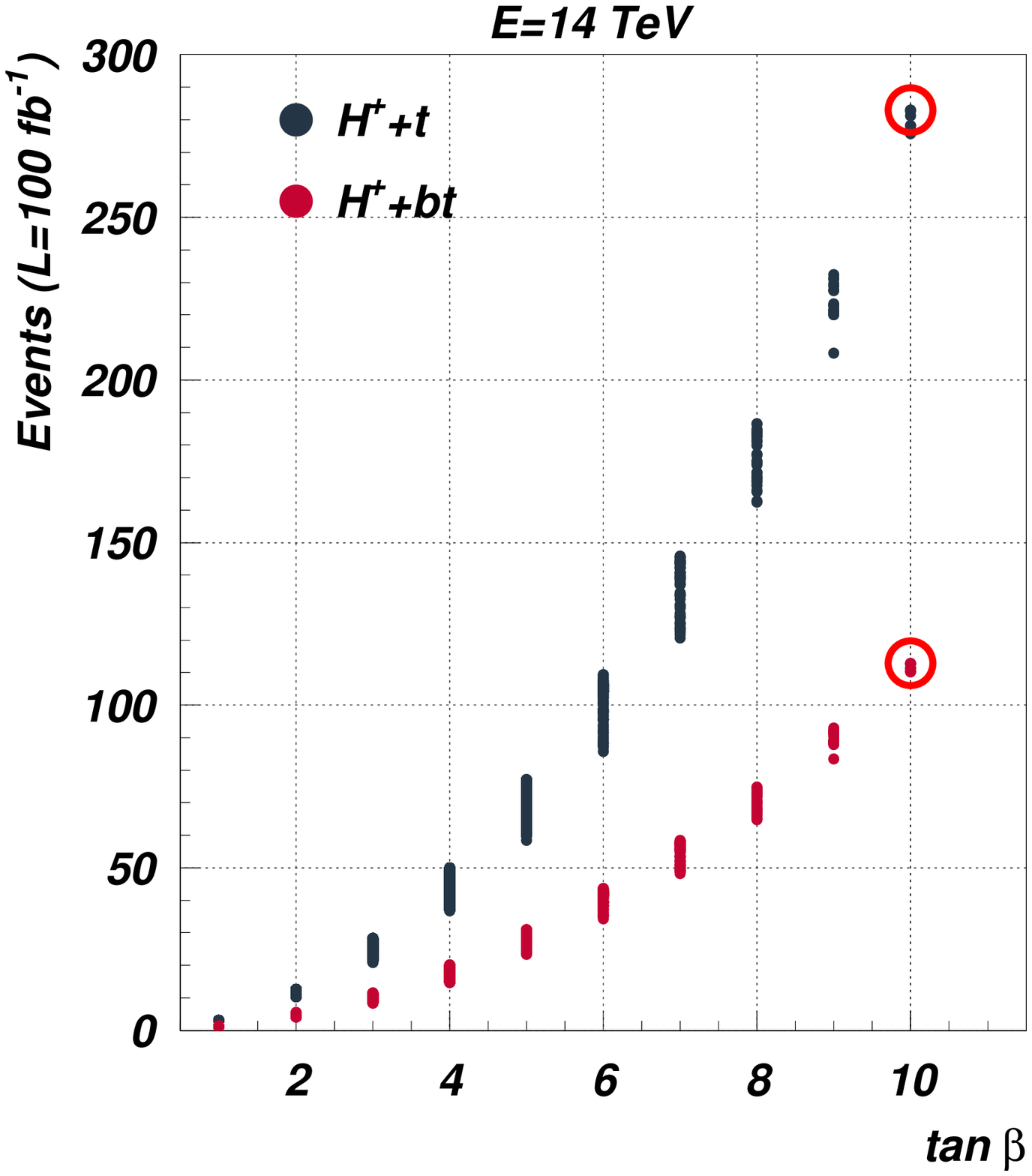}
\includegraphics[width=0.48\linewidth]{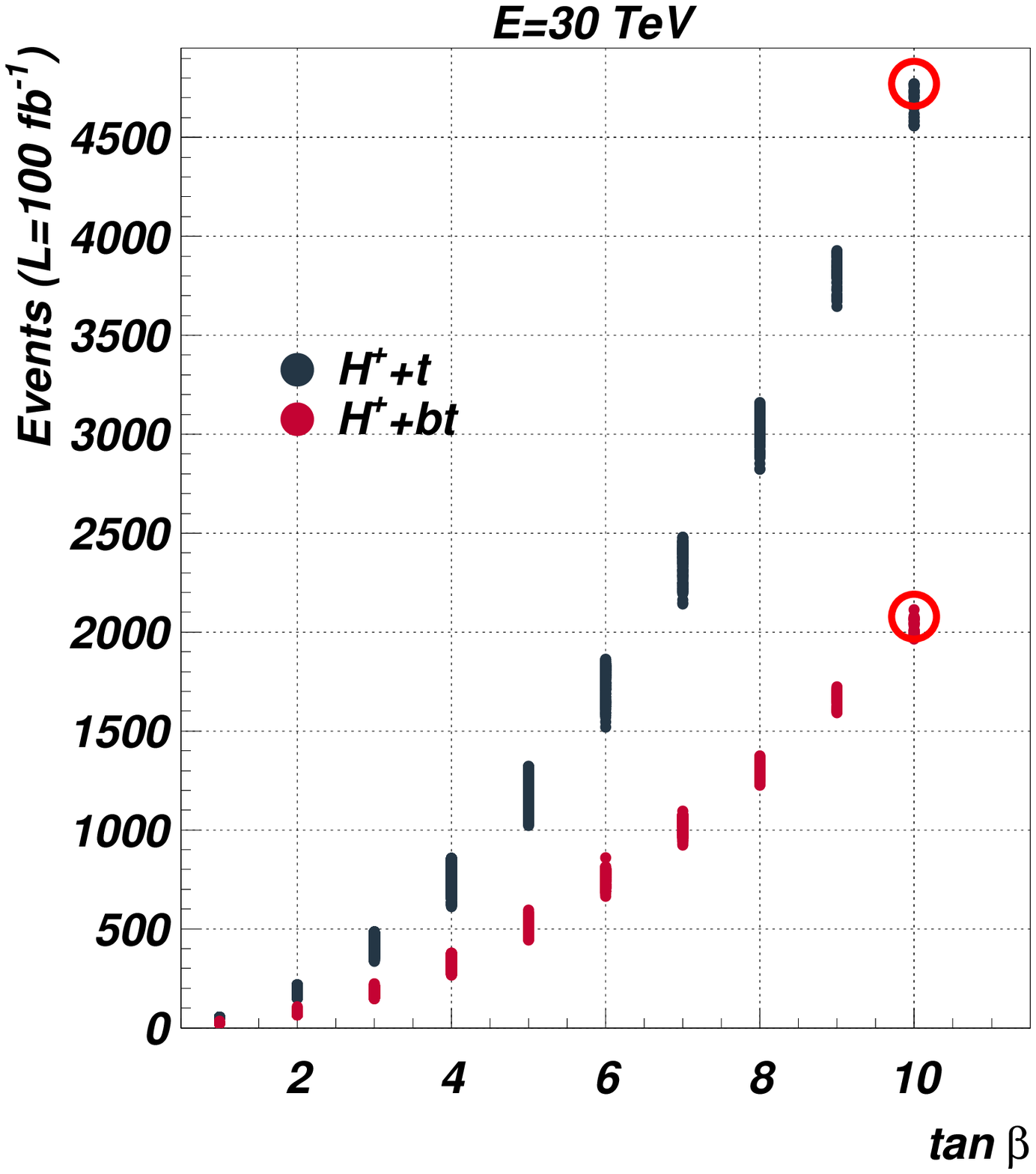}
\includegraphics[width=0.48\linewidth]{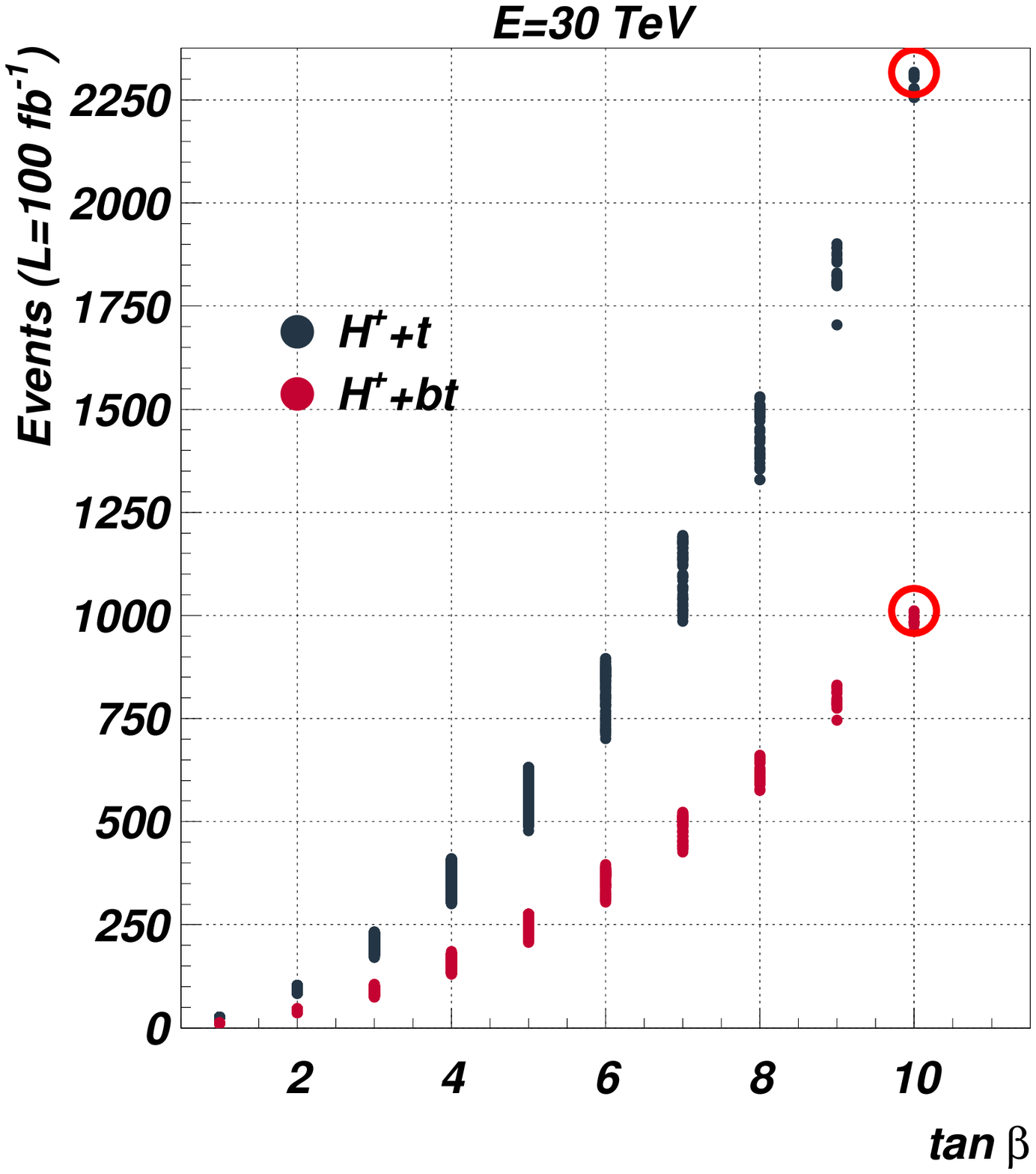}
\caption{Number of events produced via
  $pp\to H^\pm t(b)\to \tau\nu t(b)$ vs $\tan\beta$ at
  $\sqrt{s}=14$ (30)~TeV for various benchmarks with
  $M_{H^\pm}=400$~GeV shown in the upper (lower) panels. All are
  for an integrated luminosity of
  $L=100$~fb$^{-1}$. Red circles
  indicate the points with the highest production
  rate. \label{fig:TB_F1}}
\end{figure}

The last channel that requires discussion is the fermionic
channel $pp\to H^\pm tX\to \tau tX$. In figure~\ref{fig:TB_F1} the
number of events for the charged-Higgs fermionic production channel
combined with a subsequent charged-Higgs tauonic decay are plotted
against $\tan\beta$, both for the ``present'' and ``future''
scenarios. Even if the trend of the fermionic production is to
decrease for high values of $\tan\beta$, the overall rates when the BRs are
 included have a monotonically growing behaviour which is basically independent of the other parameters, since such was the case for the BRs.
 This allows one to identify the best benchmarks for this channel at the highest possible $\tan\beta$, which in the present analysis is represented by the value of $10$.

Among the many benchmark points, we selected those yielding the highest rates
 for both the bosonic and the fermionic production mechanisms when the charged Higgs decays in the tauonic mode. 
The corresponding values of the CPV 2HDM type~II parameters for such points
are collected in table~\ref{tab:benchmarks}.
In the next section, the study of their discovery reach at present and future hadronic machines is presented.

\begin{table}[ht]
\begin{footnotesize}
\begin{center}
\begin{tabular}{|c|c|c|c|c|c|c|c|c|}
\hline
\ & $\alpha_1/\pi$ & $\alpha_2/\pi$ & $\alpha_3/\pi$ & $\tan\beta$ & $M_1$ (GeV) & $M_2$ (GeV) & $\mu$ (GeV) & $M_{H^\pm}$ (GeV) \\
\hline
 $P_{B400}$ & $1.42953$ & $-0.01299$ & $0.11118$ & $7$ & $125$ & $400$ & $400$ & $400$ \\
 $P_{B500}$ & $1.43129$ & $-0.01909$ & $0.18063$  & $7$ & $125$ & $500$ & $500$ & $500$ \\
\hline
 $P_{F400}$ & $1.48311$ & $-0.01026$ & $0.10666$ & $10$ & $125$ & $400$ & $400$ & $400$ \\
 $P_{F500}$ & $1.46942$ & $-0.00928$ & $0.13918$ & $10$ & $125$ & $500$ & $500$ & $500$ \\
\hline
\end{tabular}
\end{center}
\caption{CPV 2HDM type~II parameters for the benchmark points with highest
  rates. $P_{B400}$ and $P_{B500}$ represent
  benchmark points for the bosonic case, $P_{F400}$ and $P_{F500}$
   for the fermionic case. \label{tab:benchmarks}}
\end{footnotesize}
\end{table}

\section{Signal-over-background analysis}
\label{Sec:5}
\setcounter{equation}{0}

To summarise the previous section, we will study here the following production mechanisms:
\begin{itemize}
\item[(A):] $W$-associated production: $pp\to W^\mp H^\pm\to \tau jj$ + {\tt MET};
\item[(B):] fermion-associated production: $pp\to H^\pm t(b)\to \tau t (b)$ + {\tt MET};
\end{itemize}
and compare with the competing background.

Total cross sections for the $\tau \nu$ channel for the selected benchmarks are collected in table~\ref{tab:xs_sign}, together with the $H^\pm \to \tau \nu$ branching ratios.

\begin{table}[!ht]
\begin{footnotesize}
\centering
 \begin{tabular}{|c||c|c|c||c|c|c|}
 \hline
  \multirow{2}{*}{benchmark} & \multicolumn{3}{c||}{$M_{H^\pm}=400$ GeV} & \multicolumn{3}{c|}{$M_{H^\pm}=500$ GeV} \\ \cline{2-7}
    & $\sqrt{s}=14$ TeV & $\sqrt{s}=30$ TeV & BR (\%)& $\sqrt{s}=14$ TeV & $\sqrt{s}=30$ TeV & BR (\%)  \\
 \hline
 $pp\to \tau \nu W^\pm$ & 5.26 & 32.3 & 6.92 & 1.77 & 12.5 & 5.92 \\ \hline
 $pp\to \tau \nu t$     & 6.45 & 47.5 & \multirow{2}{*}{11.9} & 2.83 & 23.1 & \multirow{2}{*}{10.4}\\
 $pp\to \tau \nu tb$    & 2.57 & 20.7 &   & 1.13 & 10.1 & \\
 \hline
 \end{tabular}
 \caption{Cross sections (in fb) and Branching Ratios for $H^\pm \to \tau \nu $.}
 \label{tab:xs_sign}
\end{footnotesize}
\end{table}

\subsection{Backgrounds}
The irreducible background to process (A) consists of the $W+Nj$ processes, with the subsequent $W\to \tau \nu_\tau$ decay. We generated 3 samples, according to the number of jets ($N=2,3$) and jet production mechanism (QCD or EW). Top-mediated backgrounds include $t\overline{t}\to tj\tau\nu$ and single top $tW\to t\tau\nu$. For better modelling of the high $M_T(\tau\nu)$ tail, the full $t\tau\nu_\tau +(0,1)j$ have been simulated in the $5$-flavours scheme. At leading order, the cross sections for these processes\footnote{Generation cuts have been used to ensure convergence: $p_T^j > 10$ GeV and $|\eta _j| < 5$ $\forall j$, $\Delta R(jj)>0.1$, $M_{jj} > 10$ GeV, 
and, for the EW sample only, $M_{jj} < 180~\text{GeV}$.} are collected in table~\ref{tab:xs_bkg_A}.
Other backgrounds include $Z+$ jets. These are subdominant and very effectively reduced when a cut on missing energy is imposed. Hence, we will not consider them here.

\begin{table}[!ht]
\centering
 \begin{tabular}{|c||c|c|c|c|c|}
 \hline
  $\sqrt{s}$ & $\tau\nu jj$ (QCD) & $\tau\nu jj$ (EW) & $\tau\nu jjj$ (QCD) & $t\tau\nu$ & $tj\tau\nu$ \\ \hline
 $\sqrt{s}=14$ TeV & 1.44$\, 10^{3}$ & 25.5 & 3.11$\, 10^{3}$ & 4.5$^a$ & 56.6\\   
 $\sqrt{s}=30$ TeV & 4.44$\, 10^{3}$ & 65.3 & 10.9$\, 10^{3}$ & 21.7$^a$ & 293.1 \\ \hline
 \end{tabular}
 \caption{Cross sections (in pb) for the backgrounds. a) No cuts applied.}
 \label{tab:xs_bkg_A}
\end{table}

For signal (B), the irreducible backgrounds are the single top and $t\overline{t}$ processes described above. Other backgrounds are the $W+Nj$ ($N\geq 3$) and $Z +$ jets. As above, the latter background is not considered. Regarding the $W+$ jets background, we considered only the $N=3$ case. Higher jet multiplicities are more suppressed and hence less important sources.

The key point to suppress the background is that in all cases in which the only source of {\tt MET} is the $\nu_\tau$ produced from $W$-boson decays to the tau lepton, the transverse mass of the latter will peak at the $W$-boson mass and rapidly fall, while the signal will peak at much larger values. We employ the following definition of the transverse mass~\cite{Barger:1987du}:
\begin{equation}
M^2_T = \left( \sqrt{M^2(\text{vis})+P^2_T(\text{vis})}+\left| \mpt \right| \right) ^2
	- \left( \vec{P}_T(\text{vis}) + \mptv\right) ^2\, .
\end{equation}
For the above reason, in the following we will restrict our analysis to the semileptonic decay modes of our final states, $\tau$ $+ Nj + $ {\tt MET}. In the type (A) signal, there will be $N=2$ jets compatible with a hadronic $W$-boson, in the type (B) signal, there will be at least one $b$-jet and a total of at least $N=3$ jets compatible with a top quark.

\subsection{Event analysis}
The selection of the objects for this analysis largely overlaps between the two cases under consideration. Jets are selected if
\begin{equation}
p_T^j > 40 \mbox{ GeV} \qquad  \mbox{and} \qquad \left| \eta _j\right| < 
\left\{ \begin{array}{c} 3.0 \mbox{ (A)}\\  2.5 \mbox{ (B)} \end{array}\right.
\end{equation}
For process (B), the jets are restricted to the coverage of the tracker to allow for $b$-tagging. We employ here the CMS ``medium'' working point~\cite{Chatrchyan:2012jua}, which has an average (in $p_T$) $b$-tagging efficiency of $70\%$, a $c$-tagging efficiency of $20\%$ (flat in $p_T$) and a mistagging rate for light jets of around $1\%$.

Concerning the tau lepton, a proper modelling of its reconstruction can be done only at detector level. To effectively emulate it in this parton level study, we apply an overall selection of
\begin{equation}
p_T^\tau > 40 \mbox{ GeV} \qquad  \mbox{and} \qquad \left| \eta _\tau\right| < 2.3\, ,
\end{equation}
with an approximate (flat) tau-tagging efficiency of $25\%$~\cite{Chatrchyan:2012zz}.

Finally, objects are required to be isolated. This means requiring
\begin{equation}
\Delta R(jj) > 0.5 \qquad  \mbox{and} \qquad \Delta R(\tau j) > 0.3 \qquad \forall j\, .
\end{equation}
In the following, we discuss the two signals separately.

\subsubsection{Bosonic-associated production mode (A)}
\begin{figure}[!t]
\centering
\includegraphics[width=0.75\linewidth]{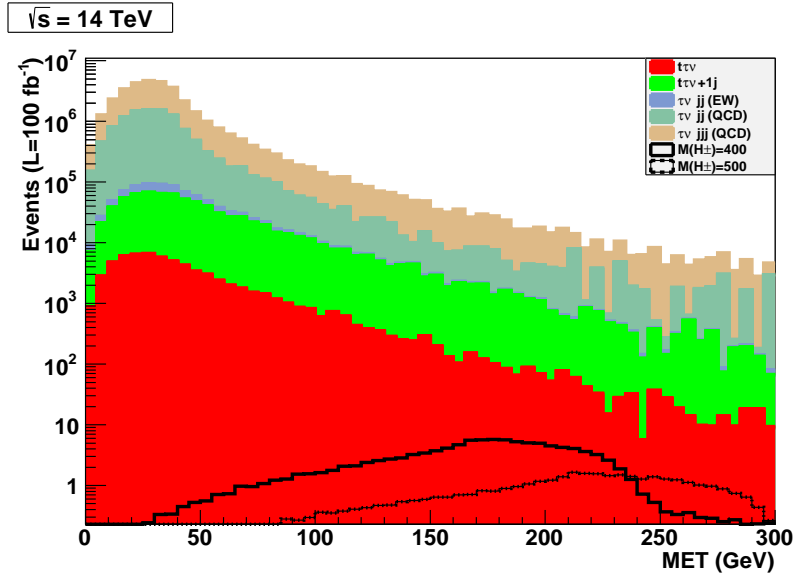}
\caption{{\tt MET} distribution for signal and background at $\sqrt{s}=14$ TeV.\label{fig:MET}}
\end{figure}

We start by presenting the analysis of the bosonic-associated production mode (A). The final state is $\tau + 2j + $ {\tt MET}. 
Its selection suffers from a complication, the way that the experiments can trigger on it. Monojet and dijet triggers require much heavier jets. We base our study on the CMS detector, that has a tau+{\tt MET} trigger, as employed in the charged-Higgs search in the tau decay mode at $\sqrt{s}=8$ TeV~\cite{CMS:2014cdp}. This trigger requires {\tt MET}~$>~70$ GeV, $p^\tau_T~>~35$ GeV, and $|\eta_\tau|<2.1$ to be fully efficient. It is however  going to be replaced for Run 2 due to the more involved experimental conditions. Trigger prototypes seem to converge to a selection of {\tt MET}~$>~200$ GeV, $p^\tau_T > 60$ GeV, and $|\eta_\tau|<2.1$ for full efficiency~\cite{tautrigger}. For the signal the {\tt MET} is expected to be much larger than for the background, since $M_{H^\pm} > M_W$ (see figure~\ref{fig:MET}). Therefore, these trigger requirements act as desired to enhance the signal over the background, and we adopt them here. However, the {\tt MET} selection is particularly severe for the $M_{H^\pm}=400$ GeV case, removing most of the events. We however want to point out that this is a parton level study only, and that jet fragmentation typically increase the overall {\tt MET}. 

Furthermore, in Ref.~\cite{CMS:2014cdp} it was pointed out that experimentally, the ratio $R_\tau = p^{\text{charged hadron}}/p^{\tau_h} > 0.7$ is used to suppress backgrounds with $W\to \tau\nu$. As explained therein, this variable is based on the helicity correlations arising from the opposite polarisation states of the $\tau$ leptons originating from the $W$ boson and the charged Higgs boson. We cannot apply the same selection here due to the lack of a simulation of tau decays. Hence, our results should be considered as conservative.

The event selection is as follows. On top of the trigger requirements for {\tt MET} and tau leptons, we require the presence of exactly 1 tau lepton and of exactly $N=2$ jets. This defines our baseline selection. Furthermore, the 2 jets in the signal are coming from a $W$-boson. We then select events that pass the following cut:
\begin{eqnarray} \label{cut:MW}
\left| M_{jj} - M_W\right| < 30 \mbox{ GeV.}
\end{eqnarray}
The cut-flow and relative efficiencies are collected in tables~\ref{Tab:eff_bos_sig} and \ref{Tab:eff_bos_bkg} for the signal and the background, respectively.

\begin{table}[h]
\begin{footnotesize}
\begin{center}
\begin{tabular}{|c||c|c|c|c||c|c|c|c|} \hline
 & \multicolumn{4}{c||}{$\sqrt{s}=14$ TeV} & \multicolumn{4}{c|}{$\sqrt{s}=30$ TeV}\\ \cline{2-9}
 & $400$ GeV & $\varepsilon (\%)$ & $500$ GeV & $\varepsilon (\%)$ & $400$ GeV & $\varepsilon (\%)$ & $500$ GeV & $\varepsilon (\%)$\\ \hline
no cuts     			      & 526  & $-$   & 177       & $-$  & 3.2\,$10^3$ & $-$   &1.2\,$10^3$& $-$  \\
baseline                              & 3.6  & 0.7   & 3.1       & 1.7  & 23.0        & 0.7   & 19.7      & 1.6   \\
$\left| M_{jj} - M_W\right| < 30$ GeV & 3.6  & 99.6  & 3.0       & 98.4 & 22.8        & 99.3  & 19.5      & 99.1\\\hline
$350 < M_T(\tau\nu)/$GeV$ < 420$         & 2.7  & 74.9  &  -        & -    & 16.1        & 70.8  &  -        & -    \\
$450 < M_T(\tau\nu)/$GeV$ < 520$         & -    & -     & 2.0       & 60.3 & -           & -     & 12.9      & 56.8 \\
\hline
\end{tabular}
\end{center}
\caption{Events and efficiencies at the LHC for the signal at
  $\sqrt{s}=14$ TeV and $\sqrt{s}=30$ TeV, for $100$ fb$^{-1}$ for
  process (A) after the application of cuts (efficiency always with
  respect to previous item). The baseline selection includes also object
  selection efficiencies. \label{Tab:eff_bos_sig}}
\end{footnotesize}
\end{table}

\begin{table}[h]
\begin{footnotesize}
\begin{center}
\scalebox{0.85}{
\begin{tabular}{|c||c|c||c|c||c|c||c|c||c|c|} \hline
$\sqrt{s}=14$ TeV & $t\tau\nu$ & $\varepsilon (\%)$ & $tj\tau\nu$ & $\varepsilon (\%)$ & $\tau\nu jj (QCD)$ & $\varepsilon (\%)$ & $\tau\nu jj (EW)$ & $\varepsilon (\%)$ & $\tau\nu jjj (QCD)$ & $\varepsilon (\%)$\\ \hline
gen. cuts     			      &450\,$10^3$& $-$   &5.7\,$10^6$& $-$  & 144\,$10^6$ & $-$  & 2.6\,$10^6$ & $-$  & 3.1\,$10^8$ & $-$ \\
baseline                              & 239       & 0.05  &2.2\,$10^3$& 0.04 & 23\,$10^3$  & 0.02 & 144       & 0.006& 49\,$10^3$ & 0.02 \\
$\left| M_{jj} - M_W\right| < 30$ GeV & 69.4      & 29.1  & 572       & 25.6 & 1.9\,$10^3$ & 8.2  & 115       & 79.9 & 5.1\,$10^3$& 10.5\\\hline
$350 < M_T(\tau\nu)/$GeV$ < 420$         &$<10^{-2}$ &$<0.01$& 0.44      & 0.08 & 28.0        & 1.5  & 2.6         & 0.2  & 20.1       & 0.4 \\
$450 < M_T(\tau\nu)/$GeV$ < 520$         &$<10^{-2}$ &$<0.01$& 0.25      & 0.04 & 17.8        & 0.9  & 2.1         & 0.2  & 10.6       & 0.2\\
\hline \hline
$\sqrt{s}=30$ TeV & $t\tau\nu$ & $\varepsilon (\%)$ & $tj\tau\nu$ & $\varepsilon (\%)$ & $\tau\nu jj (QCD)$ & $\varepsilon (\%)$ & $\tau\nu jj (EW)$ & $\varepsilon (\%)$ & $\tau\nu jjj (QCD)$ & $\varepsilon (\%)$\\ \hline
gen. cuts     			      & 2.2\,$10^6$ & $-$   & 29\,$10^6$& $-$  & 444\,$10^6$ & $-$  & 6.5\,$10^6$ & $-$ & 11\,$10^8$ & $-$\\
baseline                              & 2\,$10^3$   & 0.09  & 22\,$10^3$& 0.07  & 96\,$10^3$  & 0.02 & 387 & 0.006 & 2.2\,$10^5$ & 0.02 \\
$\left| M_{jj} - M_W\right| < 30$ GeV & 541         & 25.4  &5.6\,$10^3$& 25.7 & 6.3\,$10^3$ & 6.5  & 321 & 83.1 & 19\,$10^3$ & 8.7 \\\hline
$350 < M_T(\tau\nu)/$GeV$ < 420$         & 2.8         & 0.5   & 3.6       & 0.06 & 81.7        & 1.3  & 8.5         & 2.7  & 79.7       & 0.4 \\
$450 < M_T(\tau\nu)/$GeV$ < 520$         & 1.6         & 0.3   & 2.4       & 0.04 & 54.0        & 0.9  & 7.7         & 2.4  & 34.0       & 0.2\\
\hline
\end{tabular}}
\end{center}
\caption{Similar to table~\ref{Tab:eff_bos_sig}, but for the backgrounds. 
\label{Tab:eff_bos_bkg}}
\end{footnotesize}
\end{table}

\begin{figure}[!t]
\subfloat[]{
\includegraphics[width=0.48\linewidth]{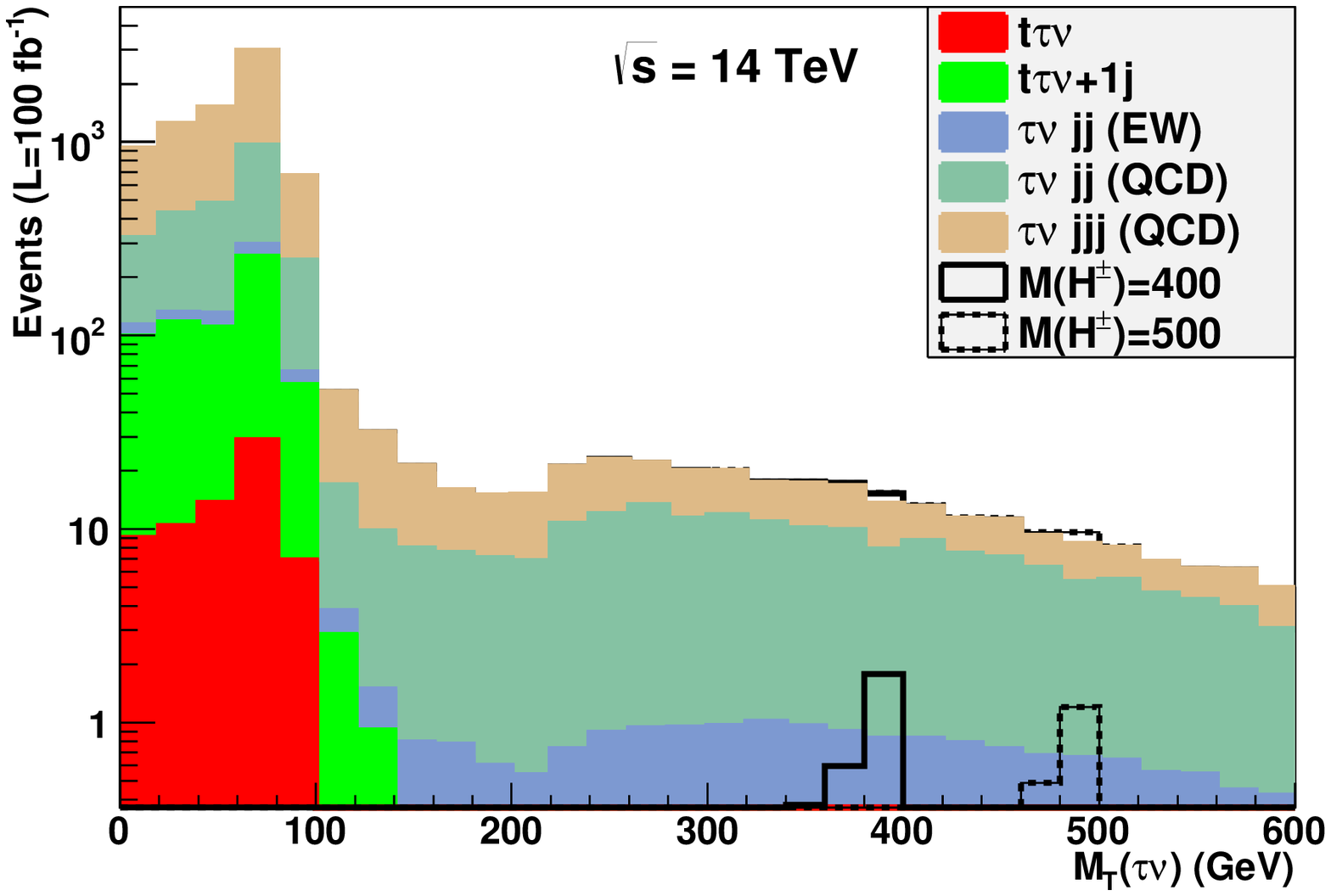}}
\subfloat[]{
\includegraphics[width=0.48\linewidth]{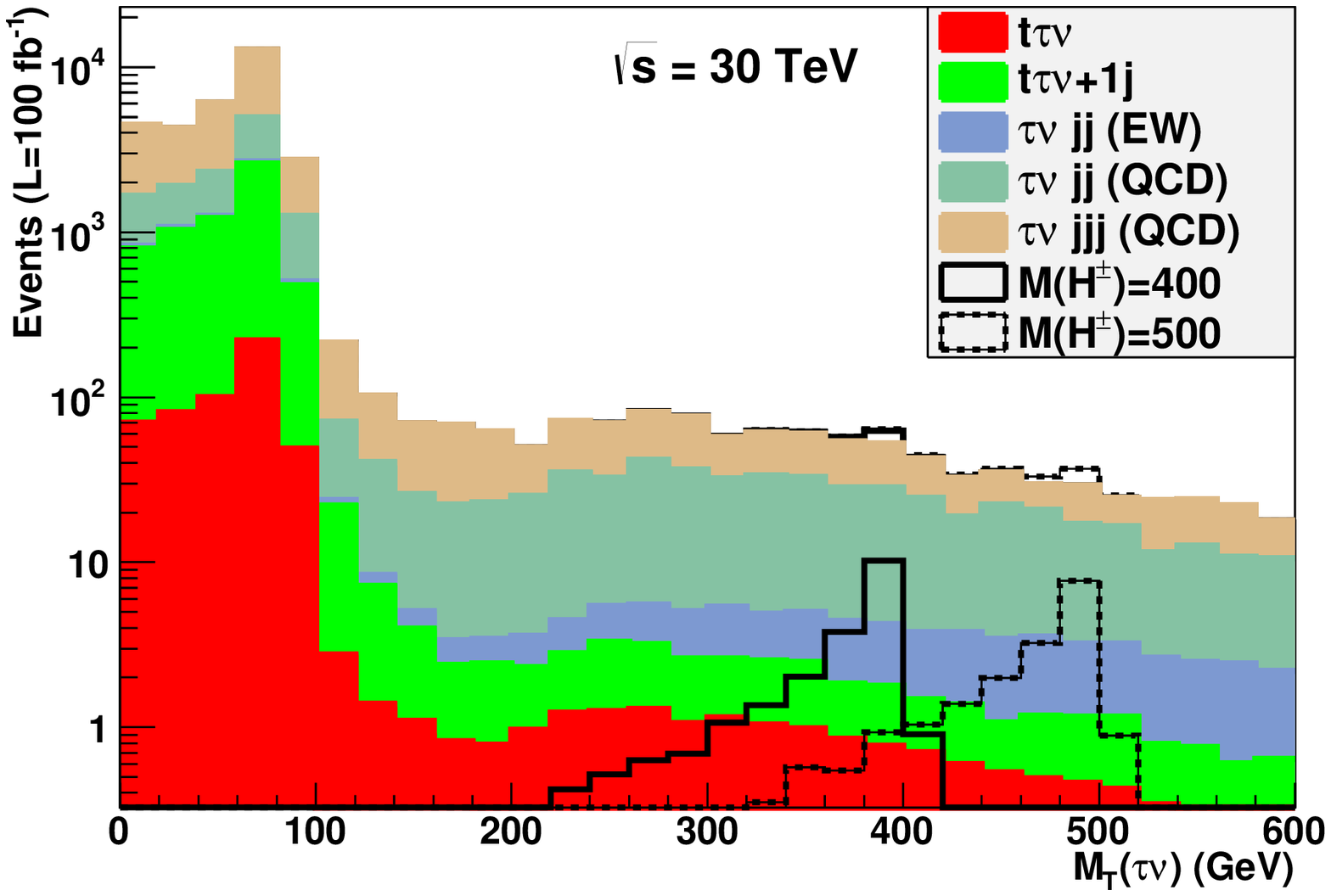}}
\caption{Transverse-mass distribution of the tau lepton at (a) $\sqrt{s}=14$ TeV and at (b) $\sqrt{s}=30$ TeV, for $100$~fb$^{-1}$ of integrated luminosity, after the application of the cut of Eq.~(\ref{cut:MW}). For reference, the signal is shown both stacked onto the background and superimposed on it. \label{fig:plot_bosonic}}
\end{figure}

If on the one hand the $H_3$-mediated production of the charged Higgs in the signal increases the production cross section, on the other hand it means that the two jets arising from the $W$-boson decays will be a bit more boosted than for the background. This is reflected in a lower efficiency to get exactly 2 isolated jets. The spectrum of the tau transverse mass is shown in figure~\ref{fig:plot_bosonic} after applying all cuts. This variable should peak at the charged Higgs mass. However, the result of the cuts previously described is not sufficient to isolate the signal from the background neither at $\sqrt{s}=14$ TeV nor at $\sqrt{s}=30$ TeV, for $100$~fb$^{-1}$ of integrated luminosity. To quantify this, we select windows around the peaks
\begin{eqnarray}
350 < &M_T(\tau\nu)/\mbox{GeV}& < 420\,, \label{cut:bosonic_peak_400} \\
450 < &M_T(\tau\nu)/\mbox{GeV}& < 520\,. \label{cut:bosonic_peak_500}
\end{eqnarray}
The relative signal-over-background significance, defined as $S/\sqrt{S+B}$, is 0.4 (1.16)~$\sigma$ and 0.35 (1.21)~$\sigma$ at $\sqrt{s}=14$ (30) TeV for the two signal benchmarks, respectively. Given that the significance in the above simplified formulation scales with $\sqrt{L}$, we expect that a $3\sigma$ observation may be possible with $\mathcal{O}(600)$ fb$^{-1}$ in the ``future'' scenario. The increase in the centre-of-mass energy is therefore argued to be a better option to assess this channel, since even the ultimate $3000$ fb$^{-1}$ of integrated luminosity option for the LHC at $\sqrt{s}=14$ TeV would merely be able to start probing the model at the 2$\sigma$ level.

\subsubsection{Fermionic-associated production mode (B)}
\begin{figure}[!t]
\subfloat[]{
\includegraphics[width=0.48\linewidth]{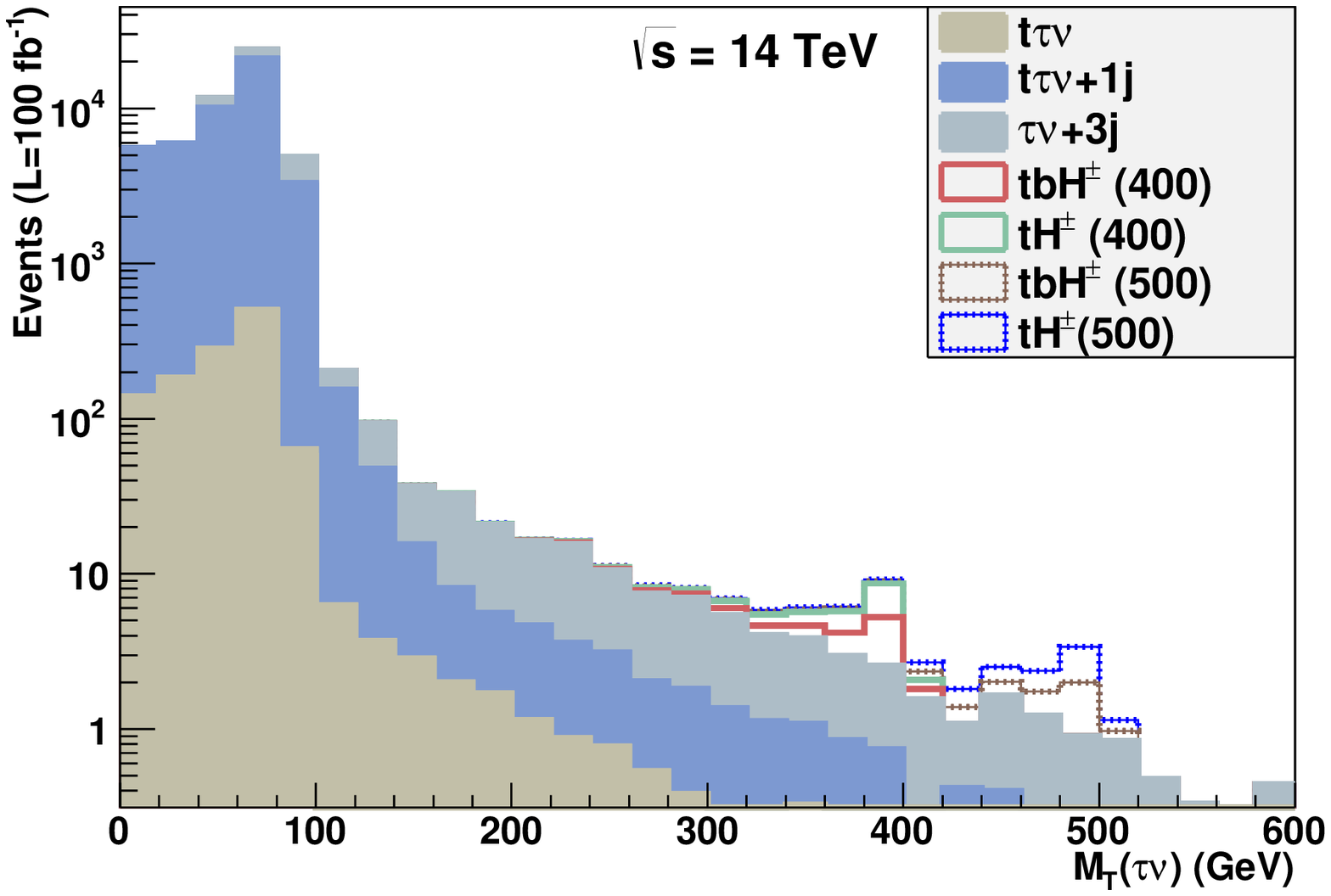}}
\subfloat[]{
\includegraphics[width=0.48\linewidth]{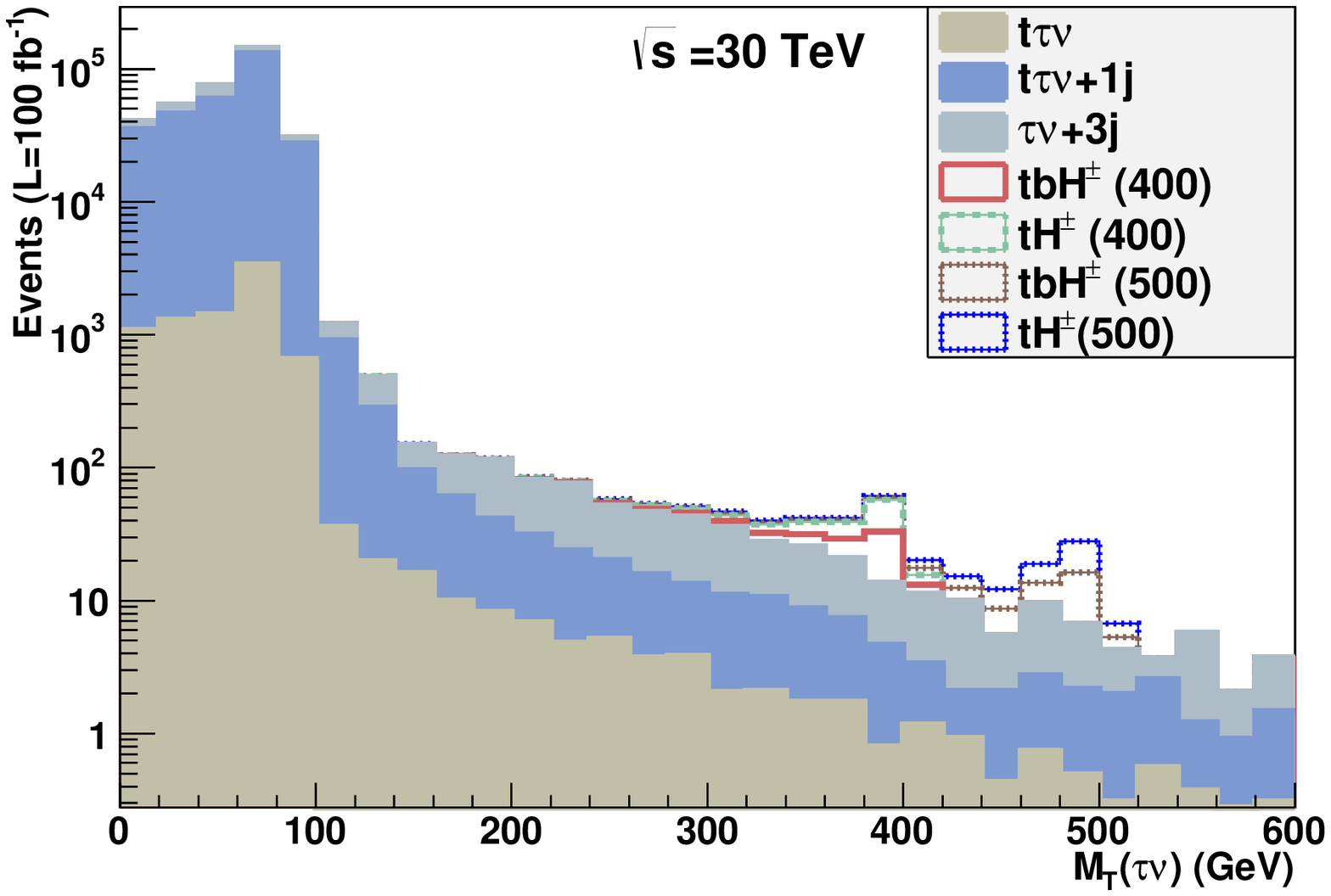}}
\caption{Transverse mass of the tau lepton for process (B) at (a) $\sqrt{s}=14$ TeV and at (b) $\sqrt{s}=30$ TeV, for $100$~fb$^{-1}$ of integrated luminosity, after the application of the cuts of Eqs.~(\ref{cut:MET_B})--(\ref{cut:Mtop}). The signal is shown stacked onto the background. \label{fig:plot_fermionic}}
\end{figure}

We now move on to the description of the fermionic production mechanism (B).
This channels suffers of no issue with triggers. Concerning the event selection, we require the presence of exactly 1 tau lepton and of at least 3 jets, of which at least one is tagged as a $b$-jet. Like for mode (A), the {\tt MET} is expected to be much larger than for the background. Furthermore, 3 jets in the signal are coming from a top quark\footnote{We did not include the $b$-tagged jet in the reconstruction of the top quark. This is because the $b$-tagged jet in the $\tau\nu tb$ production mechanisms in (B) not always comes from the top decay, unlike for $\tau\nu t$. The two signals are then analysed in the same way and can therefore be summed.}. We therefore select events that pass the cut of Eq.~(\ref{cut:MW}) and the following requirements:
\begin{eqnarray} \label{cut:MET_B}
\mbox{ \tt{MET}} &>& 100 \mbox{ GeV,} \\ \label{cut:Mtop}
\left| M_{jjj} - m_t\right| &<& 30 \mbox{ GeV.}
\end{eqnarray}
At this point, the signal is already visible on top of the background, as can be seen in figure~\ref{fig:plot_fermionic}. The cut-flow and relative efficiencies are collected in tables~\ref{Tab:eff_fermionic_signal} and \ref{Tab:eff_fermionic_bck}. We notice that the efficiency of selecting at least $3$ jets is smaller for $pp \to tH^\pm$ than for $pp \to tbH^\pm$. This is because in the latter case, 4 partons are produced and losing one jet in their selection does not alter the rate. On the contrary, in the former case only 3 partons are produced and not reconstructing one will let the event be rejected. Notice also that the jets are a bit more boosted for the signal than for the backgrounds (especially $t\tau\nu$), hence the higher selection efficiency for the latter.

\begin{table}
\begin{footnotesize}
\begin{center}
\begin{tabular}{|c||c|c||c|c||c|c||c|c|} \hline
 & \multicolumn{8}{c|}{Signal} \\ \hline
$\sqrt{s}=14$ TeV & $tbH^\pm@400$ & $\varepsilon (\%)$ & $tH^\pm@400$ & $\varepsilon (\%)$ & $tbH^\pm@500$ & $\varepsilon (\%)$ & $tH^\pm@500$ & $\varepsilon (\%)$ \\ \hline
no cuts     			      & 257  & $-$  & 646 & $-$  & 113  & -    & 283  & -    \\
$\#\tau =1$ 			      & 61.4 & 23.9 & 152.9 & 23.7 & 27.0 & 23.9 & 68.1 & 24.1 \\
$\#j \geq 3$    		      & 11.4 & 18.6 & 16.4  & 10.7 & 5.4  & 20.0 & 7.4  & 10.8 \\
$\#b \geq 1$    		      & 9.6  & 83.8 & 12.1  & 73.8 & 4.5  & 84.5 & 5.6  & 75.7 \\
MET $ > 100$ GeV                      & 8.2  & 85.9 & 10.3  & 85.2 & 4.2  & 92.2 & 5.1  & 91.6 \\
$m_t$ and $M_W$ reco.		      & 6.2  & 75.9 & 10.3  & 99.9 & 3.2  & 75.3 & 5.1  & 99.9 \\ \hline
$350 < M_T(\tau\nu)/$GeV$ < 420$  	      & 3.5  & 55.7 & 5.8   & 56.5 & -    & -    & -    & -    \\
$450 < M_T(\tau\nu)/$GeV$ < 520$  	      & -    & -    & -     & -    & 1.5  & 46.6 & 2.5  & 48.2 \\
\hline
\hline
$\sqrt{s}=30$ TeV & $tbH^\pm@400$ & $\varepsilon (\%)$ & $tH^\pm@400$ & $\varepsilon (\%)$ & $tbH^\pm@500$ & $\varepsilon (\%)$ & $tH^\pm@500$ & $\varepsilon (\%)$ \\ \hline
no cuts     			      & 2078 & $-$  & 4750 & $-$  & 1014 & -    & 2314 & -    \\
$\#\tau =1$ 			      & 473  & 22.8 & 1087 & 22.9 & 235  & 23.2 & 539  & 23.3 \\
$\#j \geq 3$   			      & 83.7 & 17.8 & 110.6& 10.2 & 44.1 & 18.8 & 57.5 & 10.7 \\
$\#b \geq 1$    		      & 70.2 & 83.8 & 83.7 & 75.7 & 37.4 & 84.8 & 44.0 & 76.6 \\
MET $ > 100$ GeV                      & 59.7 & 85.1 & 72.1 & 86.1 & 34.6 & 92.5 & 40.6 & 92.3 \\
$m_t$ and $M_W$ reco.		      & 45.1 & 75.6 & 72.1 & 99.9 & 25.4 & 73.4 & 40.6 & 99.8 \\ \hline
$350 < M_T(\tau\nu)/$GeV$ < 420$         & 24.9 & 54.9 & 40.1 & 55.7 & -    & -    & -    & -    \\
$450 < M_T(\tau\nu)/$GeV$ < 520$         & -    & -    & -     & -   & 12.4 & 48.9 & 20.0 & 49.4 \\
\hline
\end{tabular}
\end{center}
\caption{Events and efficiencies for $100$ fb$^{-1}$ for the process (B) signal after the application of cuts (efficiency always with respect to previous item), for (top) $\sqrt{s}=14$ TeV and (bottom) $\sqrt{s}=30$ TeV. Cuts 1 and 2 include also object selection efficiencies.}\label{Tab:eff_fermionic_signal}
\end{footnotesize}
\end{table}

\begin{table}
\begin{footnotesize}
\begin{center}
\begin{tabular}{|c||c|c||c|c||c|c|} \hline
 & \multicolumn{6}{c|}{Background}\\ \hline
$\sqrt{s}=14$ TeV & $\tau\nu t$ & $\varepsilon (\%)$ & $\tau\nu t j$ & $\varepsilon (\%)$ & $\tau\nu jjj$ & $\varepsilon (\%)$ \\ \hline
no cuts     			      
& 452\,$10^3$ & $-$ & 5.7\,$10^6$ & $-$ & 311\,$10^6$ & - \\
$\#\tau =1$ 			      
&68\,$10^3$  & 15.1& 709\,$10^3$ & 12.5& 22.8\,$10^6$ & 7.3 \\
$\#j \geq 3$    		      
& 10\,$10^3$  & 15.1& 463\,$10^3$ & 65.3& 1.1\,$10^6$ & 5.0 \\
$\#b \geq 1$    		      
& 7.6\,$10^3$ & 74.3& 409\,$10^3$ & 88.3 & 98.4\,$10^3$ & 8.7 \\
MET $ > 100$ GeV                      
& 1.2\,$10^3$ & 15.7& 44.8\,$10^3$& 10.9 & 6.5\,$10^3$ & 6.6 \\
$m_t$ and $M_W$ reco.		      
& 1.2\,$10^3$ & 99.9& 36.5\,$10^3$& 81.5 & 3.2\,$10^3$ & 49.5 \\ \hline
$350 < M_T(\tau\nu)/$GeV$ < 420$  	      
& 0.6 & 0.05& 1.4         & 4\,$10^{-3}$ & 2.5 & 0.08\\
$450 < M_T(\tau\nu)/$GeV$ < 520$  	      
& 0.1         & 0.01& 0.23         & 6\,$10^{-4}$ & 1.2 & 0.04\\
\hline
\hline
$\sqrt{s}=30$ TeV & $\tau\nu t$ & $\varepsilon (\%)$ & $\tau\nu t j$ & $\varepsilon (\%)$ & $\tau\nu jjj$ & $\varepsilon (\%)$ \\ \hline
no cuts     			      
& 2.2\,$10^6$& $-$ & 30\,$10^6$  & $-$ & 1.1\,$10^9$  & $-$\\
$\#\tau =1$ 			      
& 318\,$10^3$& 14.6& 3.6\,$10^6$ & 12.2 & 72\,$10^6$  & 6.6\\
$\#j \geq 3$   			      
& 49.0\,$10^3$ & 15.4& 2.3\,$10^6$ & 62.9 & 3.7\,$10^6$ & 5.2\\
$\#b \geq 1$    		      
& 36.8\,$10^3$ & 75.2& 2.0\,$10^6$ & 88.4 & 331\,$10^3$ & 8.9\\
MET $ > 100$ GeV                      
& 7937       & 21.5& 292\,$10^3$ & 14.4 & 46\,$10^3$  & 13.9\\
$m_t$ and $M_W$ reco.		      
& 7337       & 99.9& 227\,$10^3$ & 77.7 & 11\,$10^3$  & 23.7\\ \hline
$350 < M_T(\tau\nu)/$GeV$ < 420$         
& 5.9        & 0.07& 13.0        & 6\,$10^{-3}$ & 15.8 & 0.14\\
$450 < M_T(\tau\nu)/$GeV$ < 520$         
& 1.7        & 0.02& 4.2        & 2\,$10^{-3}$ & 3.4 & 0.03\\
\hline
\end{tabular}
\end{center}
\caption{Similar to Table~\ref{Tab:eff_fermionic_signal}, but for the background.}\label{Tab:eff_fermionic_bck}
\end{footnotesize}
\end{table}

To quantify the signal-over-background significance, we further select the region of interest as in (A), see Eqs.~(\ref{cut:bosonic_peak_400})--(\ref{cut:bosonic_peak_500}). It is seen that $100$ fb$^{-1}$ of integrated luminosity is not sufficient to probe the two individual channels for either value of the charged Higgs mass at the ``present'' LHC configuration. The combination of the channels scores $2.5\sigma$ and $1.7\sigma$ for $M_{H^\pm}=400$ and $500$ GeV, respectively. In turn, 3 (5) sigma discovery can be achieved with 150 (400) and 320 (900)~fb$^{-1}$ for the two benchmarks. In the ``future'' configuration instead, $100$ fb$^{-1}$ of integrated luminosity is sufficient for the discovery of the combined signals for both masses, reaching $6.5\sigma$ and $5.0\sigma$, respectively. The individual channels (in the same order as in table~\ref{tab:xs_sign}) can be probed at $5\sigma$ with $90$ and $180$ fb$^{-1}$ for $M_{H^\pm}=400$ GeV, and with $160$ and $300$ fb$^{-1}$ for $M_{H^\pm}=500$ GeV.

This production mechanism certainly proves to be the best to access the tauonic decay mode of the charged Higgs. This channel could already be discovered at the LHC Run~2 for the benchmark points here considered. Its low yield, on the other hand, implies that it is very hard to exclude it experimentally. If no signal is observed, it is argued that the increase in centre-of-mass energy will certainly be a better option than the increase in total luminosity. 

\section{Conclusions}
\label{sec:conclusions}
\setcounter{equation}{0}
We have performed scans over the parameter space of the complex 2HDM with
type~II Yukawa couplings allowing for CP violation. We do however restrict
ourselves to the case of $\lambda_6=\lambda_7=0$, in order to
constrain flavour-changing neutral currents. 
The potential is reconstructed from ``physical parameters''
\cite{Khater:2003wq}, like masses and mixing angles. The familiar
theoretical constraints are taken into account, including checking for
false vacua as discussed in Ref.~\cite{Grzadkowski:2010au}.
The amount of CP violation is very much constrained by the fact that
the $H_1ZZ$ coupling is ``SM-like'' \cite{Grzadkowski:2014ada}, but also by the
constraint from the electron EDM
\cite{Regan:2002ta,Pilaftsis:2002fe,Barr:1990vd}.

We studied in detail the production of a charged Higgs boson, distinguishing the ``bosonic'' (i.e. $pp\to H^\pm W^\mp$) from the ``fermionic'' (i.e. $pp\to H^\pm t(b)$) channels. 
The update of our previous investigation of the bosonic channel with the subsequent $H^\pm \to W^\pm H_1 \to W^\pm \overline{b}b $ decay chain confirmed that this channel has still a large scope at the LHC Run 2. We then focused on the
often-discussed tauonic decay mode ($H^\pm \to \tau \nu_\tau$), and analysed its production cross sections in both channels. The possibility of a resonant bosonic production via $H_3$ largely increases its expected rates, even above the fermionic one. Furthermore, the resonant production allowed us to neglect the box contributions in the evaluation of the bosonic cross sections. In Appendix~\ref{App:A} is is shown that this approximation is especially justified when the bosonic channel gets resonant.

The comparison to the backgrounds in the subsequent signal-over-background analysis showed however that the fermionic channel is still the preferred one for analysis. It can yield a discoverable rate of events already at the LHC Run 2 (although rather challenging), that can be definitely established either in the high luminosity option or if an upgrade in centre-of-mass energy is pursued.
The bosonic production mode instead can be probed only at colliders with higher 
centre-of-mass energies, although large integrated luminosities are still required.

\vspace*{10mm} {\bf Acknowledgements.}  
LB has received support from the Theorie-LHC France initiative 
of the CNRS/IN2P3 and by the French ANR 12 JS05 002 01 BATS@LHC.
The work of GMP has been supported by the European Community's Seventh
Framework Programme (FP7/2007-2013) under grant agreement n.~290605
(COFUND: PSI-FELLOW) and by the Swiss National Science Foundation (SNF) under 
contract 200021-160156. PO has been supported by the Norwegian Research Council.

\appendix
\section{Box contribution to the $pp\to H^\pm W^\mp$ process}
\label{App:A}
\setcounter{equation}{0}

\begin{figure}[h]
\centering
\includegraphics[]{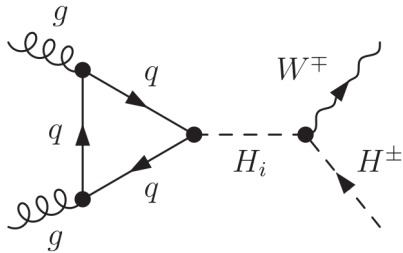}
\caption{Vertex-type diagram ($\triangle$). Here $i=1..3$ is the Higgs boson mass eigenstate index.\label{fig:vertex}}
\end{figure}

\begin{figure}[h]
\includegraphics[]{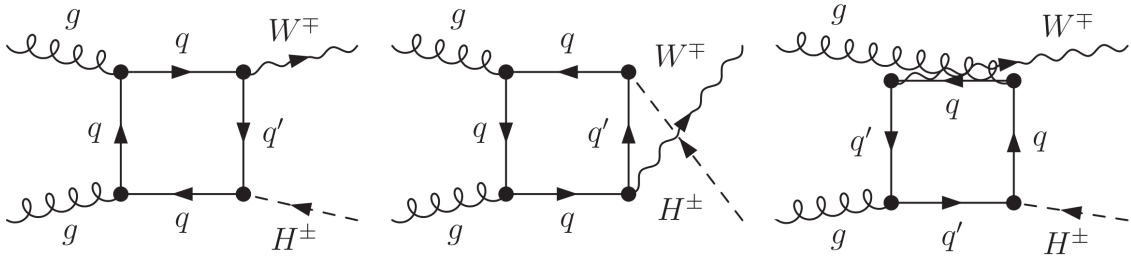}
\caption{Box-type diagrams ($\Box$).\label{fig:boxes}}
\end{figure}

In this appendix we comment on the approximation used throughout this work, i.e. we neglect the box diagrams in the computation of the ``bosonic'' signal cross sections at the LHC. In figure~\ref{fig:vertex} we display the topology of amplitudes used to evaluate the ``bosonic'' signal cross section at leading order. In figure~\ref{fig:boxes} the topology for the box amplitudes are shown. Notice here that these amplitudes are only schematic, a summation over all intermediate states, as well as the sum of the Hermitian conjugated amplitudes, has to be performed in the complete computation.

\begin{figure}[h]
\includegraphics[width=0.75\linewidth]{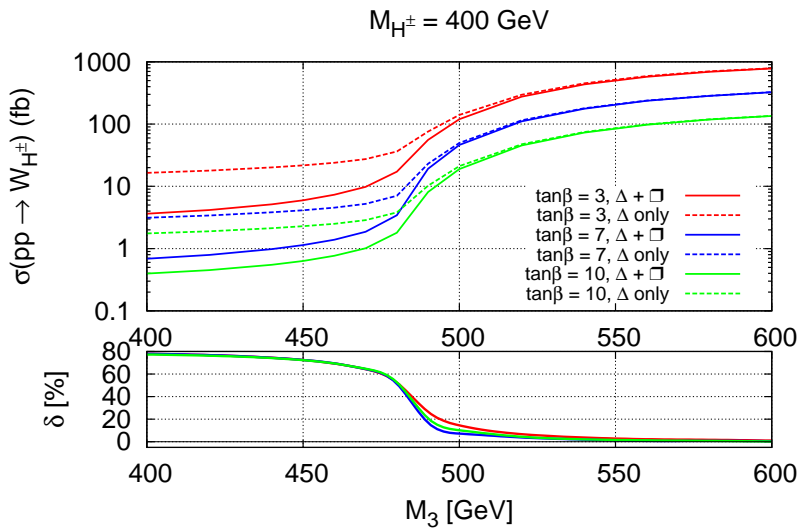}
\caption{Relative impact of box diagrams on the cross section evaluation. The $\tan\beta = 7$ case corresponds to benchmark $P_{B400}$ in table~\ref{tab:benchmarks}. Cross sections are for the LHC at $\sqrt{s}=14$ TeV.\label{fig:plotbox}}
\end{figure}

Total rates were already computed in the literature at leading order for the 2HDM~\cite{Asakawa:2005nx}, and for the (N)MSSM beyond the leading order (see e.g. Refs.~\cite{Dao:2010nu,Enberg:2011ae}), where an effective Born approximation was devised. On the contrary, we decided to compare the cross section at the LHC for $\sqrt{s}=14$ TeV, when only triangle topologies are considered and when, in addition to the latter, also box diagrams are included. The net effect of including box topologies is a reduction of the total cross section, due to negative interference. The cross sections are shown in figure~\ref{fig:plotbox} in the upper frame, while in the lower frame we quantify the discrepancy of our approximation,
\begin{equation}
\delta = 1-\frac{\sigma_{\triangle + \Box}}{\sigma_\triangle}\, ,
\end{equation}
as a function of the mass of the heaviest Higgs boson, $H_3$. For the
sake of the computation, the latter mass has been varied artificially from its physical value
while keeping all other parameters fixed, recomputing the boson width
each time. Then, we computed the cross sections for each value.

Figure~\ref{fig:plotbox} clearly shows that as the process mediated by
an $s$-channel $H_3$ boson gets resonantly enhanced, the approximation
of neglecting the box diagrams is more and more valid. For smaller
masses, the approximation does not hold, but such values are not
interesting since they are not physical. We collect
comparison figures evaluated at the physical $M_3$ value (consistent
with the other input parameters) for a few
$\tan\beta$ values in table~\ref{tab:box_importance}. 

\begin{table}[ht]
\begin{center}
\begin{tabular}{|c|c|c|}
\hline
 ~$\tan\beta$~ & ~$M_3$ (GeV)~ & $\delta$  \\
\hline
 $3$  & $517.7$ & ~$7.0~\%$~ \\
 $7$  & $507.3$ & ~$5.3~\%$~ \\
 $10$ & $510.9$ & ~$5.8~\%$~ \\
\hline
\end{tabular}
\end{center}
\caption{Relative importance of neglecting the box diagrams at the physical $H_3$ mass values. \label{tab:box_importance}}
\end{table}

We quantify the effect of neglecting the box diagrams in this work in an $\mathcal{O}(10\%)$ difference as compared to the correct cross section evaluation. This is compatible with the parton level accuracy of our study. Hence, our approximation is justified.

\bibliographystyle{JHEP}
\bibliography{biblio}

\end {document}